%% file: note1532.tex
\documentclass[11pt]{article}
\usepackage{graphicx}
\usepackage[normalem]{ulem}
\usepackage{rotating,epsfig,hhline,amsmath,feynmp}
\usepackage{lscape,dcolumn}
\usepackage{cite}
\newcommand{\BABARPubYear}    {06}

\newcommand{\BABARConfNumber} {004}
\newcommand{\SLACPubNumber} {11948}

\input pubboard/babarsym
 
\long\def\inst#1{\par\nobreak\kern 4pt\nobreak
    {\it #1}\par\vskip 10pt plus 3pt minus 3pt}

\def\pcm {\ensuremath{p^*}\xspace}
\def\Lc  {\ensuremath{\Lambda_c^+}\xspace}
\def\Dsp {\ensuremath{D^+_{(s)}}\xspace}

\def\llp {\ensuremath{\ellp\ell'^-}\xspace}
\def\DsTopiphi  {\ensuremath{\Ds\to\pip\phi_{KK}}\xspace}
\def\DcTopiphi  {\ensuremath{\Dp\to\pip\phi_{KK}}\xspace}
\def\DspTopiphi  {\ensuremath{\Dsp\to\pip\phi_{KK}}\xspace}
\def\LctopKpi    {\ensuremath{\Lc\to p\Km\pip}\xspace}

\def\DtoPiee {\ensuremath{\Dp\to\pip\epem}\xspace}

\def\DstoPimm {\ensuremath{\Ds\to\pip\mumu}\xspace}

\def\DtoKmm {\ensuremath{\Dp\to\Kp\mumu}\xspace}

\newcolumntype{d}[0]{D{.}{.}{-1}}
\newcolumntype{p}[0]{D{p}{\pm}{-1}}

\begin{document}
{\pagestyle{empty}
                                                                                
\begin{flushright}
\babar-CONF-\BABARPubYear/\BABARConfNumber \\
SLAC-PUB-\SLACPubNumber \\
July 2006 \\
\end{flushright}
                                                                                
\par\vskip 5cm

\begin{center}
{\Large \bf Search for Flavor-Changing Neutral-Current Charm Decays}
\end{center}
\bigskip

\begin{center}
\large The \babar\ Collaboration\\
\mbox{ }\\
\today
\end{center}
\bigskip \bigskip

\begin{center}
\large \bf Abstract
\end{center}
We search for rare FCNC charm decays of the form $X_c^+\to h^+\llp$,
where $X_c^+$ is a charm hadron, $h$ is a pion, kaon or proton, and
$\ell^{(}{'}^{)}$ is an electron or a muon. In the pion and kaon
modes, we study both \Dp and \Ds decays, while in the proton modes we
study \Lc decays.  Based on a data sample of 288\invfb of \epem
collisions collected by \babar, we set preliminary 90\% confidence
level limits between 4 to 40$\times10^{-6}$ for the branching
fractions of the different decay modes. For most decay modes, our
analysis provides a significant improvement over previous results.

\vfill
\begin{center}
                                                                                
Submitted to the 33$^{\rm rd}$ International Conference on High-Energy Physics,
ICHEP 06,\\
26 July---2 August 2006, Moscow, Russia.
                                                                                
\end{center}
                                                                                
\vspace{1.0cm}
\begin{center}
{\em Stanford Linear Accelerator Center, Stanford University,
Stanford, CA 94309} \\ \vspace{0.1cm}\hrule\vspace{0.1cm}
Work supported in part by Department of Energy contract DE-AC03-76SF00515.
\end{center}
                                                                                
\newpage
} 

\input  pubboard/authors_ICHEP2006.tex                                                                           
\input introduction

\input data

\input selection

\input fitshapes

\input systematics

\input fits

\input result

\section{ACKNOWLEDGMENTS}
\label{sec:Acknowledgments}

\input pubboard/acknowledgements.tex

\end{document}

%% file: pubboard/authors_ICHEP2006.tex
\begin{center}
\small

The \babar\ Collaboration,
\bigskip

%
{B.~Aubert,}
{R.~Barate,}
{M.~Bona,}
{D.~Boutigny,}
{F.~Couderc,}
{Y.~Karyotakis,}
{J.~P.~Lees,}
{V.~Poireau,}
{V.~Tisserand,}
{A.~Zghiche}
\inst{Laboratoire de Physique des Particules, IN2P3/CNRS et Universit\'e de Savoie,
 F-74941 Annecy-Le-Vieux, France }
% ...

\end{center}\newpage

%% file: introduction.tex
\section{INTRODUCTION}

In the Standard Model (SM), flavor-changing neutral-current (FCNC)
processes cannot occur at the tree level. FCNC processes therefore provide
an excellent tool for investigating the quantum corrections in the SM
as a way to search for evidence of physics beyond the SM. FCNC
processes have been studied extensively for $K$ and $B$ mesons in
$\Kz-\Kzb$ and $\Bz-\Bzb$ mixing processes and in rare FCNC decays,
such as $s\to d\ellell$, $b\to s\gamma$ and $b\to s\ellell$ decays.
The present measurements of these processes agree with SM predictions \cite{thurth},
but there are strong ongoing efforts to improve both the measurements
and the theoretical predictions, and to measure new effects, such as
CP violation, in FCNC processes.

\begin{figure}
\begin{fmffile}{ctoullWbox} 
  \fmfframe(20,40)(10,40){  
   \begin{fmfgraph*}(180,100)
    \fmfstraight
    \fmfleft{i0,i1,i2,i3,i4,i5,i6,i7}
    \fmfright{o0,o1,od,od2,od3,o2,od4,o3}
    \fmftop{t1,t2,t3,t4}
    \fmflabel{$c$}{i1}
    \fmflabel{$u$}{o1}
    \fmflabel{$\ell$}{o2}
    \fmflabel{$\ell$}{o3}
    \fmf{fermion}{i1,v1,v2,o1}
    \fmffreeze
    \fmf{boson,label=$W$}{v1,v3}
    \fmf{boson,label=$W$}{v2,v4}
    \fmf{phantom}{v3,m3,t2}
    \fmf{phantom}{v4,m4,t3}
    \fmffreeze
    \fmf{fermion}{o2,v4,v3,o3}
   \end{fmfgraph*} 
  }
\end{fmffile}
\begin{fmffile}{ctoullPeng}  
  \fmfframe(10,40)(1,40){ 
   \begin{fmfgraph*}(180,100) 
    \fmfstraight
    \fmfleft{i0,i1,i2,i5,i6,i7} 
    \fmfright{o0,o1,od,o2,o3,od4} 
    \fmftop{t1,t2,t3,t4}
    \fmflabel{$c$}{i1} 
    \fmflabel{$u$}{o1}
    \fmflabel{$\ell$}{o2}
    \fmflabel{$\ell$}{o3}
    \fmf{fermion}{i1,v1}
    \fmf{plain}{v1,v3}
    \fmf{fermion}{v3,v4}
    \fmf{plain}{v4,v2}
    \fmf{fermion}{v2,o1} 
    \fmf{boson,right,label=$W$}{v1,v2}
    \fmffreeze
    \fmf{boson,label=$\gamma/Z^0$,label.side=left}{v4,v5}
    \fmf{fermion}{o3,v5,o2}
    \fmf{phantom}{t3,v5}
   \end{fmfgraph*} 
  }
\end{fmffile}

\caption{\label{fig:Feynman} Standard model short-distance
contributions to the $c\to u\ellell$ transition.}
\end{figure}
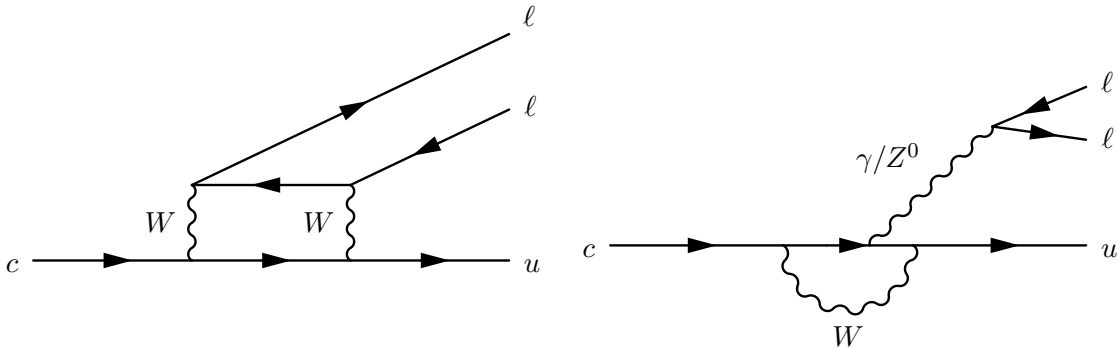

FCNC processes in the charm sector have received less attention and
the experimental upper limits are currently above the SM predictions.
In the SM very small signals are expected, as a consequence of
effective Glashow-Iliopoulos-Maiani (GIM) cancellation. For instance,
the $c\to u\ellell$ transitions illustrated in Fig.~\ref{fig:Feynman}
lead to branching fractions for $D\to X_u\ellell$ of $O(10^{-8})$
\cite{Burdman,Fajfer}. This contribution is masked by the presence of
long-distance contributions from intermediate vector resonances such
as $D\to X_u V, V\to\ellell$. These are predicted to have branching
fractions of $O(10^{-6})$ \cite{Burdman,Fajfer}. In $c\to u\ellell$
transitions, the effect of these resonances can be separated from
short-distance contributions by studying the invariant mass of the
$\ellell$ pair. In radiative charm decays, $c\to u\gamma$, the
long-distance contributions make it impossible to study the underlying
short-distance physics \cite{BurdmanRadiative}.

Several extensions to the SM have been studied and their impact on
$D\to X_u\ellell$ decay rates estimated~\cite{Burdman,LittleHiggs}. The largest
possible effect is expected in R-parity violating supersymmetric
models. Depending on the size of the R-parity violating couplings,
branching fractions of up to $O(10^{-5})$ for different $D\to
X_u\ellell$ decays are possible. This is within the reach of our
experimental sensitivity of $O(10^{-6}$--$10^{-5})$, depending on the
decay mode.

There is a large group of possible FCNC charm decays to be
measured. The best existing limits are for the branching fractions
BF$(\Dz\to\ellp\ell'^-)$ \cite{BaBarD0ll} and are $O(10^{-6})$ at 90\%
CL.  In this analysis we search for FCNC charm decays of the form
$X^+_c\to h^+\ellp\ell'^-$, where the two leptons $\ellp$ and $\ell'^-$
can each be either an electron or muon.  The charge-conjugate decay
modes are implied here and throughout this note. Current upper limits
\cite{CLEOhee,Dzero,focus,fixedtarget}
range from no limits in some of
the baryon modes to $5\times10^{-6}$.  The FCNC decay combinations we examine are
$\Dp\to\pip\ellp\ell'^-$, $\Ds\to\Kp\ellp\ell'^-$ and
$\Lc\to\proton\ellp\ell'^-$. Decays where the two leptons are of
different flavor are lepton-family violating and therefore forbidden
in the SM. We also search for $\Dp\to\Kp\ellp\ell'^-$ and
$\Ds\to\pip\ellp\ell'^-$ decays, but these require both quarks in the
charm meson to change flavor.

The only long-distance contributions relevant at the current
experimental sensitivity are from $\Dp\to\pip\phi$ and
$\Ds\to\pip\phi$ decays. The branching fractions to $\pip\ellell$
through these two resonance decays are $1.8\times10^{-6}$ and
$1.1\times10^{-5}$, respectively \cite{PDG}. In this analysis, we measure
the total rate of decay excluding a region around the $\phi$ resonance
in the invariant mass of the \ellell pair.

The measured FCNC decay yields are converted into branching ratios by
normalizing them to the yields of known charm decays.  We choose
normalization modes that have kinematics similar to the FCNC decays,
so that most of the systematic effects not related to particle identification
cancel in the branching ratio. For the \Dp and \Ds decays we use
decays to $\pip\phi$ and for the \Lc we use $\Lc\to\proton\Km\pip$
decays. The measured branching fractions for these modes are listed in
Table~\ref{tab:normBF}. The $\phi$ decays are reconstructed in only
the $\Kp\Km$ decay modes.  This introduces an
additional branching fraction for $\phi\to\KpKm$, which is $0.491\pm
0.006$ \cite{PDG}. We use the abbreviation \DspTopiphi to denote
the $\Dsp\to\pip\phi, \phi\to\KpKm$ decays.

\begin{table}
\caption{\label{tab:normBF} Branching fractions \cite{PDG} for
the charm decays used for normalization.}
\begin{center}
\begin{tabular}{lp}
Decay mode                &  \multicolumn{1}{r}{Branching Fraction}  \\ \hline
$\Dp\to\pip\phi$          & (6.2 p 0.6)\times10^{-3} \\ 
$\Ds\to\pip\phi$          & (36  p 9)\times10^{-3} \\ 
$\Lc\to\proton\Km\pip$    & (50  p 13)\times10^{-3}  \\ \hline
\end{tabular}
\end{center}
\end{table}

%% file: data.tex
\section{THE \babar\ DETECTOR AND DATASET}
\label{sec:babar}

The measurements are performed using data collected by the \babar{}
detector \cite{ref:babar} at the \pep2\ storage ring at SLAC. The data
sample used comprises an integrated luminosity of 263\invfb collected
from \epem collisions at the \FourS resonance and 25\invfb collected
40\mev below the \FourS resonance.  For event simulation we use the
Monte Carlo (MC) generator EVTGEN \cite{evtgen} with a full detector
simulation based on GEANT4 \cite{geant4}.  Signal and the \LctopKpi MC
events are generated with a 3-body phase-space distribution while
\DspTopiphi MC events are generated with a Breit-Wigner for the $\phi$
decay. All signal events are simulated as \ccbar continuum events.
Samples of simulated generic \ccbar and $uds$ continuum events and \BB
decays corresponding to 1.4--5 times the recorded data sample are used
to study background contributions.

%% file: selection.tex
\section{ANALYSIS METHOD}

Initial charm hadron candidates are formed from one track identified
as either a pion, kaon or proton and two tracks each of which is identified as
an electron or a muon. Typical electron (muon) identification efficiency
is about 94\% (60\%). Hadron identification efficiencies average about
98\%, 87\% and 80\% for pions, kaons and protons, respectively. The
two tracks identified as leptons are required to have opposite
charge. For candidates with a pion or kaon track, the invariant mass
of the charm candidate is required to be between 1.7 and 2.1\gevcc,
while for the candidates with protons it is required to be between 2.2
and 2.4\gevcc.  Charm hadrons from continuum production typically
carry most of the energy of the initial quarks. We therefore require
the momentum, \pcm, of the charm hadron candidate in the \epem center-of-mass frame to be
larger than 2.8 to 3.5\gevc depending on decay mode. This removes a
large fraction of the background. Charm hadrons produced
in \B decays are kinematically limited to be below about 2.2 \gevc and
are therefore not used in this analysis.

After the initial event selection, significant combinatoric background
contributions remain from semileptonic \B decays and low-multiplicity
QED events.  These background sources have been studied using events
from invariant mass sidebands in data and from MC samples. The final event
selection criteria are chosen based on these studies to minimize the
expected upper limit on the $X_c^+\to h^+\llp$ branching ratio under
the assumption that no signal is present.

The QED events are mainly radiative Bhabha, initial state radiation
and two-photon events.  These are easily identified by their low
multiplicity and/or highly jet-like structure. They are suppressed by
requiring at least five tracks in an event and a minimum event
sphericity \cite{bjorken} of 0.13 calculated in the \epem
center-of-mass frame.

We suppress the background from semileptonic $B$ decays by
rejecting events with evidence of neutrinos and requiring the two
leptons to come from a common point.  The latter is achieved by a
tight requirement on the probability of the vertex fit $\chi^2$
($P(\chi^2)>0.05$) and on the distance of closest approach between the
two leptons ($<250\mum$).

\begin{figure}
\centerline{\includegraphics[width=12cm]{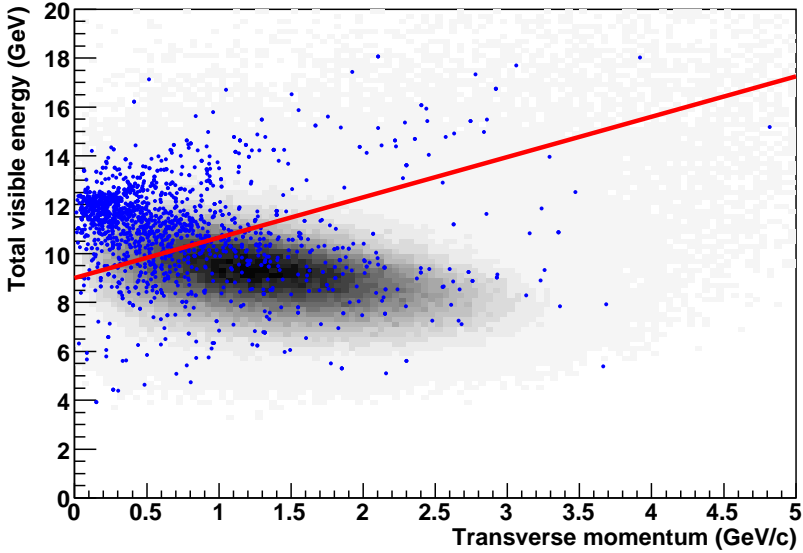}}
\caption{\label{fig:missingE} Total energy vs net transverse momentum
for $\Dp\to\pip\epem$ MC signal events (points) and background events
from generic \BpBm events (shaded histogram) for the \pip\epem
mode. The events below the red line are rejected in the final event
selection. }
\end{figure}

The neutrinos are not directly detected and therefore show up as
missing energy in the event. This is measured in two ways.  We first
calculate the total energy in an event from all reconstructed neutral
clusters and tracks in an event, assuming the pion mass for
tracks. Second we calculate the net transverse momentum in an event
(with respect to the beam axis) by adding the momentum vectors of all
neutral clusters and tracks. The neutral clusters are assumed to be photons.
The more the transverse momentum deviates from zero the more likely it
is to be a semileptonic $B$ decay.  Figure~\ref{fig:missingE} shows
the distributions of the two variables for $\Dp\to\pip\epem$ signal MC
events and for $\pip\epem$ candidates in generic MC \BpBm events.
Selecting on a linear combination of the two variables as indicated in
the figure, 74\% of the signal events are kept while 84\% of the \BpBm
events are rejected.

In the \epem decay modes, there is a significant contribution at low
$\epem$ mass from $\piz$ decays. Some of these are from decays directly to
$\epem(\gamma)$ while others are $\piz\to\gamma\gamma$, where one of the photons
converts to \epem. These are removed together with $\gamma$ conversions by
requiring $m(\epem)>200\mevcc$.

For $\Ds\to\pip\llp$ decays, we furthermore require the presence of a
photon candidate with an energy above 100\mev and
$|m(h\llp\gamma)-m(h\llp)-0.1438\mevcc|<0.015\mevcc$. This selects \Ds
mesons coming from $\Dss\to\Ds\gamma$ decays and helps reduce the non-\Ds
background. 

  The requirement on the momentum \pcm of the charm hadron is decay mode
dependent:
\begin{itemize}
\item For $\Dp\to h\llp$, $\pcm>3.3\gevc$.
\item For $\Ds\to h\llp$, $\pcm>3.5\gevc$.
\item For $\Lc\to p\llp$, $\pcm>2.8\gevc$.
\end{itemize}

 For the $\Dsp\to\pip\ellell$ decay modes, we exclude events with
$0.95<m(\epem)<1.05\gevcc$ and $0.99<m(\mumu)<1.05\gevcc$. The
excluded regions for the two decay modes are different due to the
larger radiative tails in the $\pip\epem$ decay mode. In order to make
further cross-checks, we also select a sample with $\Dsp\to\pi\phi,
\phi\to\ellell$ candidates. For this sample we require
$0.995\gevcc<m(\epem)<1.030\gevcc$, $1.005\gevcc<m(\mumu)<1.030\gevcc$
and $\pcm>3.1\gevc$.

For the normalization decay modes the same selection is applied except
for the lepton particle identification requirements and the
restriction on the distance of closest approach between the two lepton
candidates. Instead the kaons and pions in the \DspTopiphi and
\LctopKpi decay modes are required to be specifically identified. For
the \DspTopiphi decay modes, we further require the invariant mass of
the kaon pair to be within 15\mevcc of the world-average value for the
$\phi$ mass \cite{PDG}. Since the selection of $\Ds\to\pip\llp$ decays
requires the presence of a photon from $\Dss\to\Ds\gamma$ decays whereas
the $\Ds\to\Kp\llp$ selection does not, the selection of $\DsTopiphi$
decays is performed separately for the two decay modes.

%% file: fitshapes.tex
The invariant mass distributions of the $h^+\llp$ candidates are fitted
using an extended unbinned maximum likelihood fit. The probability
density function (PDF) for signal events is given by the
so-called Crystal Ball function \cite{CBPaper} in order to account for
radiative tails:
\begin{equation}
  P_{CB}(m;\mu,\sigma,\alpha,n)=\left\{\begin{array}{ll}
                   e^{-0.5(\frac{m-\mu}{\sigma})^2} & \mbox{if }m\ge \mu-\alpha\sigma.\\
                   e^{-0.5\alpha^2}\left(\frac{n\sigma}{n\sigma-\alpha^2\sigma-\alpha(m-\mu)}\right)^n & \mbox{if } m< \mu-\alpha\sigma.
                   \end{array}\right.
\end{equation}
The four parameters, $\mu,\sigma,\alpha$ and $n$, are obtained from
fits to signal MC and kept fixed during fits to the data, leaving only
the overall normalization as a free parameter.  The fitted width of
the Gaussian component ($\sigma$) is found to lie between 6 and
9\mevcc depending on the decay mode. The MC events have been corrected
to reproduce particle identification efficiencies measured in various
control modes. The yields from the MC fits are used to calculate the
signal efficiencies, which range between 0.3 and 5.3\%. The modes with
low efficiency are decays with two muons and decays requiring a photon
from $\Dss$ decays. The muon identification efficiency is very low for
muons with momentum below 1\gevc.

For the normalization decay modes, the radiative effects are
negligible and we use the sum of two Gaussian distributions with a
common mean to describe the \Dp, \Ds and \Lc signals. All parameters
are free in the fits to data.
 
The invariant mass distribution of the combinatoric background events
for the signal modes are described by first-order polynomials.  For
the normalization decay modes, a second-order polynomial is used. The
background parameters are allowed to vary freely in all cases.

An additional background comes from hadronic charm decays where two
hadrons are misidentified as leptons. Pions have a probability of
about 2\% (0.1\%) to be identified as a muon (electron) depending
on the pion momentum and angle. This background component is
negligible in the signal modes with electrons and is therefore only
included in decay modes with two muons. The shape of this background
is obtained from MC samples of hadronic three-body charm decays. Each
event is weighted according to the probability of misidentifying a
pion as a muon. The misidentification probability is measured from
data using samples of $\Dz\to\Km\pip$ decays. The misidentified
hadronic charm decays are reconstructed at slightly lower $h^+\mumu$
mass than the signal events. The peak mass is shifted by about
15\mevcc which is sufficient separation for the yield to be determined
by the likelihood fit instead of relying on the MC prediction.

 For the charm meson modes two signals are fitted simultaneously since
the \Dp and \Ds mesons can decay to the same final state. However,
since the optimal event selection criteria are different for the \Dp
and \Ds modes, only one of the fitted yields is used from each fit.

For most modes we do not see a significant signal; in addition to
determining a central value for the yields, we set upper limits on the
branching ratios at 90\% confidence level (CL). This is chosen as the
point where the negative log likelihood is $1.355$ above its minimum
value. In order to always have a physical (positive) upper limit, we
only consider the minimum at or above 0 signal events. This generally
results in conservative confidence intervals. Systematic uncertainties
from the signal efficiency and the normalization mode, detailed in the
next section, are included as a Gaussian constraint in the likelihood
expression.

%% file: systematics.tex
\section{SYSTEMATIC STUDIES}

\label{sec:systematics}

Most systematic effects are expected to cancel in the branching ratio
since they affect the signal and normalization modes equally. We
therefore only have to account for differences in selection,
acceptance and decay kinematics.  Table~\ref{tab:Systematics} gives a
summary of all the systematic uncertainties related to the branching
ratio calculation. An additional systematic is assigned for the
estimation of the signal yield. The details of the different
uncertainties are given below.

\begin{table}
\caption{\label{tab:Systematics} Summary of the multiplicative systematic uncertainties for all the decay modes.
DOCA is the uncertainty due to the distance of closest approach requirement and PID is the uncertainty from
the particle identification.}
\begin{center}
\begin{tabular}{lcccccc}
           & Normalization &          &      &     &      &        \\
Decay mode &       Mode    & MC stat. & DOCA & PID & \pcm & Total  \\ \hline
\input tables/FullSystematics.tex
\end{tabular}
\end{center}
\end{table}

Systematic uncertainties related to the signal PDF parameters obtained
from MC are investigated in two ways. First, the PDF parameters for
data and MC are compared in the fits to the normalization
modes. Differences can be due either to general data-MC tracking
differences or to uncertainty in the PDG mass used in the
simulation. Second, fits to the invariant mass of $\jpsi\to\ellell$
candidates from inclusive \B decays are compared between data and
MC. The second comparison is sensitive to effects associated with
lepton reconstruction. Based on these studies, fits with the mean mass
shifted by up to 2.5\mevcc, depending on the decay mode, are
performed. Based on the same studies, the widths of the signal PDFs
are changed by $\pm 3\%$. The fit giving the highest upper limit on
the branching ratio is used for the final result.

For the background shape assumption, the signal fits are repeated
using a second order polynomial as the background PDF instead of the
nominal first order polynomial. The result of the fit with the higher
limit is used to quote the upper limit, unless variations of the
signal PDF yield an even higher limit.

In the normalization modes, the statistical uncertainties from the
fits, the MC statistics and uncertainties from the signal and
background shapes are all at or below 1\%. The main uncertainty
related to the normalization modes comes from how the efficiency is
affected by sub-resonances in the decays. This is estimated to be about 3.5\% for
each of the \Dsp modes and 1.1\% for the \LctopKpi mode.

The requirement that the distance of closest approach between the two
leptons be below 250\mum is not applied to the normalization mode and
any MC-data difference can therefore not be expected to cancel in the
branching ratio. Studies of the $\jpsi\to\ellell$ samples show efficiency
differences between data and MC to be less than 0.3\%.

The efficiency of the particle identification  has associated
systematic uncertainties. We assign 0.6\% for each pion, 1.1\% for
each kaon, 1\% for each electron and 2\% for each muon. We do not apply a
systematic uncertainty for the protons, since both the signal and the
normalization mode contain a proton and the uncertainty therefore
cancels.  Uncertainties from the same types of particles are added
linearly, while for different types they are added in quadrature.

The efficiencies of the signal and normalization modes do not have the
same dependence on \pcm. Varying the \pcm distribution used in the MC
to better match the data changes the ratio of signal and normalization
mode efficiencies by less than 2\% for all decay modes.

In the calculation of the signal efficiency we assume that the decays
follow a three-body phase-space model. The selection efficiency has
some dependence on where the decay lies in the Dalitz plane, so this
assumption introduces a systematic uncertainty.  Ignoring the regions
we explicitly remove in the selection and the very high end of the
$m(\llp)$ distribution, the efficiency typically varies less than 25\%
around the average as a function of $m(\llp)$. This model dependence
is not included in the systematic uncertainty.

%% file: tables/FullSystematics.tex
$\Dp\to\pip\epem$ & 3.7\% & 1.0\% & 0.3\% &   3\% & 2\% &  5.3\% \\
$\Dp\to\pip\mumu$ & 3.7\% & 1.8\% & 0.3\% &   5\% & 2\% &  6.8\% \\
$\Dp\to\pip\ep\mun$ & 3.7\% & 1.3\% & 0.3\% &   3\% & 2\% &  5.3\% \\
$\Dp\to\pip\mup\en$ & 3.7\% & 1.3\% & 0.3\% &   3\% & 2\% &  5.3\% \\
$\Ds\to\pip\epem$ & 4.0\% & 1.8\% & 0.3\% &   3\% & 2\% &  5.7\% \\
$\Ds\to\pip\mumu$ & 4.0\% & 3.3\% & 0.3\% &   5\% & 2\% &  7.5\% \\
$\Ds\to\pip\ep\mun$ & 4.0\% & 2.3\% & 0.3\% &   3\% & 2\% &  5.9\% \\
$\Ds\to\pip\mup\en$ & 4.0\% & 2.3\% & 0.3\% &   3\% & 2\% &  5.9\% \\
$\Dp\to\Kp\epem$ & 3.7\% & 1.1\% & 0.3\% &   4\% & 2\% &  5.9\% \\
$\Dp\to\Kp\mumu$ & 3.7\% & 2.2\% & 0.3\% &   5\% & 2\% &  6.9\% \\
$\Dp\to\Kp\ep\mun$ & 3.7\% & 1.5\% & 0.3\% &   4\% & 2\% &  6.0\% \\
$\Dp\to\Kp\mup\en$ & 3.7\% & 1.5\% & 0.3\% &   4\% & 2\% &  6.0\% \\
$\Ds\to\Kp\epem$ & 3.7\% & 1.1\% & 0.3\% &   4\% & 2\% &  5.9\% \\
$\Ds\to\Kp\mumu$ & 3.7\% & 2.3\% & 0.3\% &   5\% & 2\% &  6.9\% \\
$\Ds\to\Kp\ep\mun$ & 3.7\% & 1.6\% & 0.3\% &   4\% & 2\% &  6.0\% \\
$\Ds\to\Kp\mup\en$ & 3.7\% & 1.6\% & 0.3\% &   5\% & 2\% &  6.7\% \\
$\Lc\to\proton\epem$ & 2.5\% & 1.0\% & 0.3\% &   2\% & 2\% &  3.9\% \\
$\Lc\to\proton\mumu$ & 2.5\% & 2.3\% & 0.3\% &   4\% & 2\% &  5.6\% \\
$\Lc\to\proton\ep\mun$ & 2.5\% & 1.7\% & 0.3\% &   3\% & 2\% &  4.7\% \\
$\Lc\to\proton\mup\en$ & 2.5\% & 1.7\% & 0.3\% &   3\% & 2\% &  4.7\% \\
$\Dp\to\pip\phi_{\epem}$ & 3.7\% & 0.8\% & 0.3\% &   3\% & 2\% &  5.2\% \\
$\Dp\to\pip\phi_{\mumu}$ & 3.7\% & 1.7\% & 0.3\% &   5\% & 2\% &  6.8\% \\
$\Ds\to\pip\phi_{\epem}$ & 3.7\% & 0.8\% & 0.3\% &   3\% & 2\% &  5.2\% \\
$\Ds\to\pip\phi_{\mumu}$ & 3.7\% & 1.6\% & 0.3\% &   5\% & 2\% &  6.7\% \\ \hline

%% file: fits.tex
\section{PHYSICS RESULTS}

\begin{figure}[t]
\centerline{\includegraphics[width=8cm]{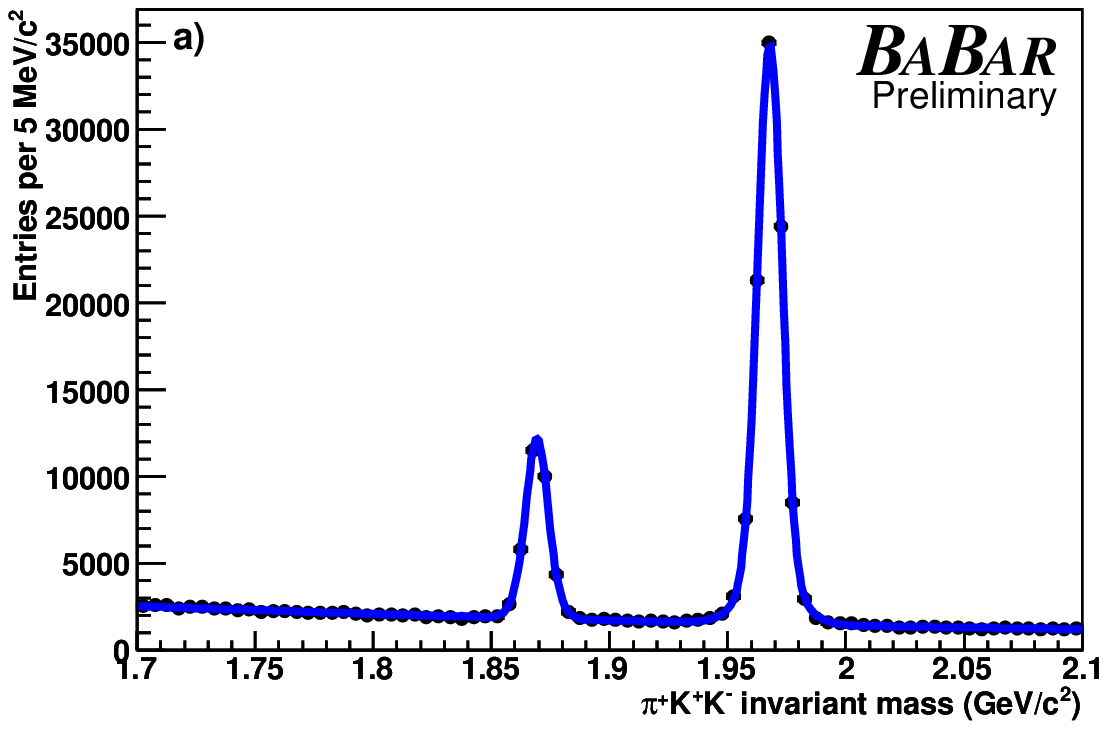}\includegraphics[width=8cm]{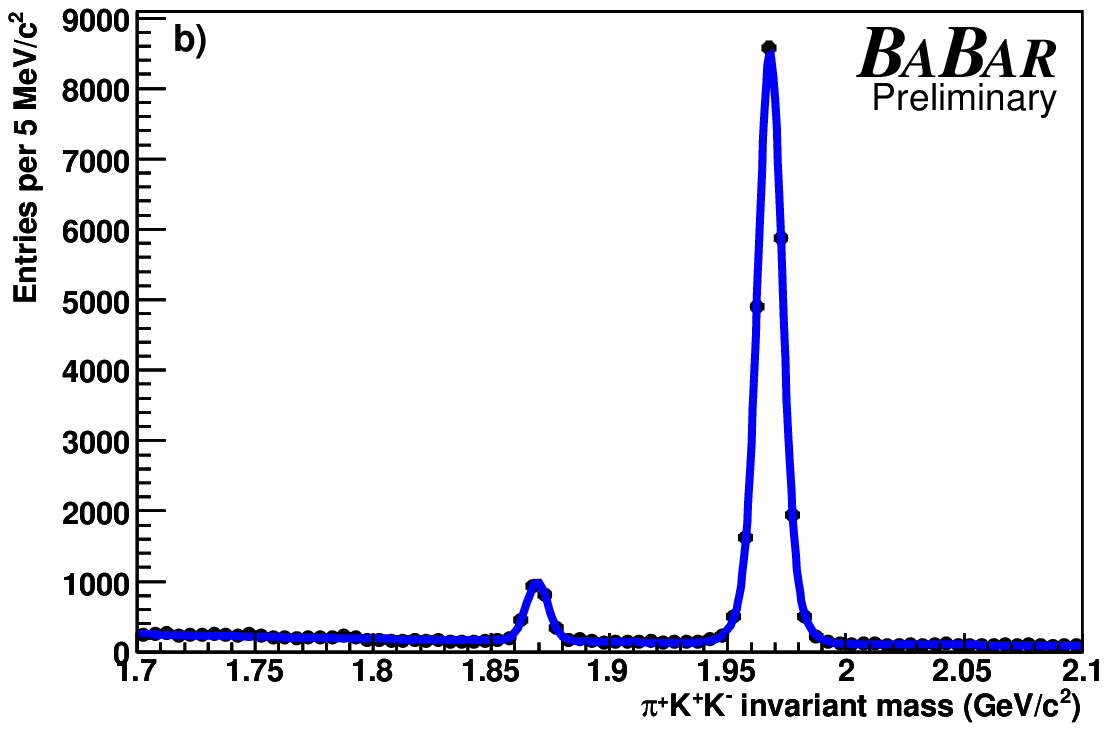}}
\centerline{\includegraphics[width=8cm]{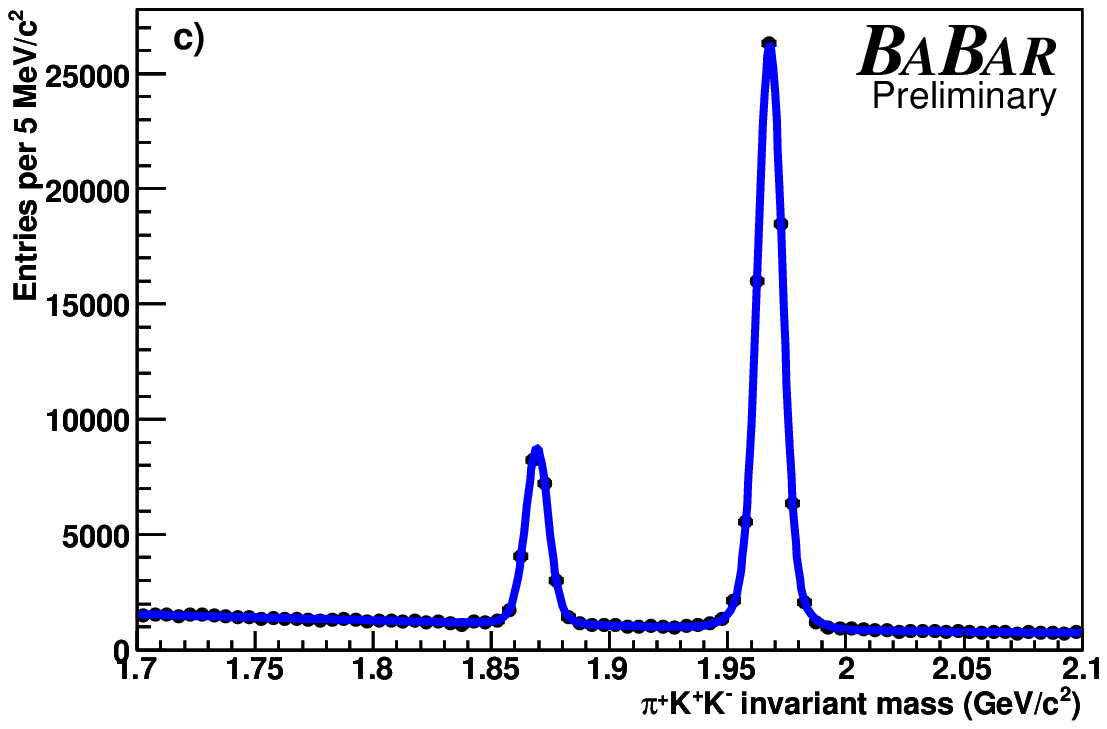}\includegraphics[width=8cm]{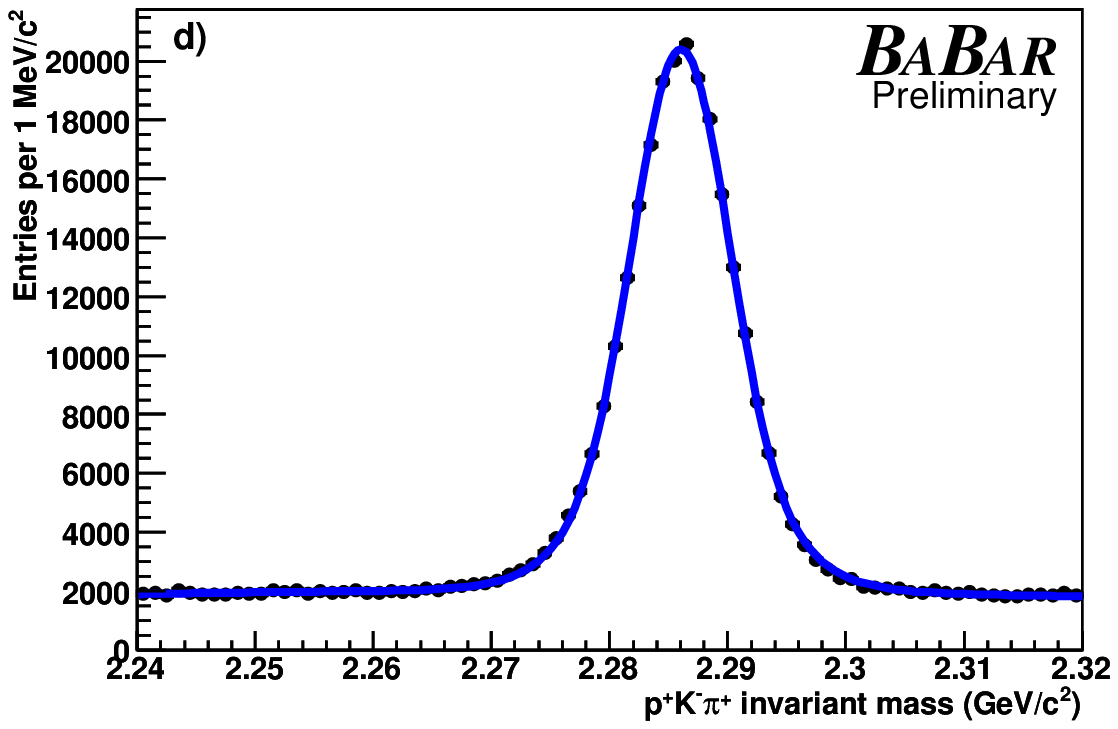}}
\caption{\label{fig:RefFit} Invariant mass distribution for
$\pip\phi_{\Kp\Km}$ candidates with a) $\Dp\to\pip\llp$ and
$\Dp\to\Kp\llp$ selection criteria, b) $\Ds\to\pip\llp$ selection
criteria and c) $\Ds\to\Kp\llp$ selection
criteria. d) Invariant mass distribution for
$\LctopKpi$ candidates with the $\Lc\to\proton\llp$ selection
criteria. The solid lines are the result of
a fit to double-Gaussian signals and a second-order polynomial for the
background.}
\end{figure}

The invariant mass distributions of the normalization decay modes and
the corresponding fits are shown in Fig.~\ref{fig:RefFit}. The
$\pip\KpKm$ invariant mass distribution has both \Dp and \Ds signals
and is shown multiple times, because the \Dp and \Ds selections have
different \pcm requirements and the $\Ds\to\pip\llp$ selection
requires a $\gamma$ from a $\Dss$ decay.  Each selection requires its
own fit in order to cancel most systematic uncertainties.  The fitted
signal yields are listed in Table~\ref{tab:referenceFits}. The table
also lists the efficiencies estimated from signal MC.

\begin{table}
\caption{\label{tab:referenceFits} MC efficiency and fitted yields in data for the normalization modes.
The $\pip\KpKm$ distribution is fitted after the different selection criteria matching those for the corresponding signal modes are applied.}
\begin{center}
\begin{tabular}{llpr}
Decay mode & Selection Mode  & N_{\mathrm{sig}} & Efficiency \\ \hline
\input tables/FullReferenceFits.tex
\end{tabular}
\end{center}
\end{table}

\begin{figure}[t]
\centerline{\includegraphics[width=8cm]{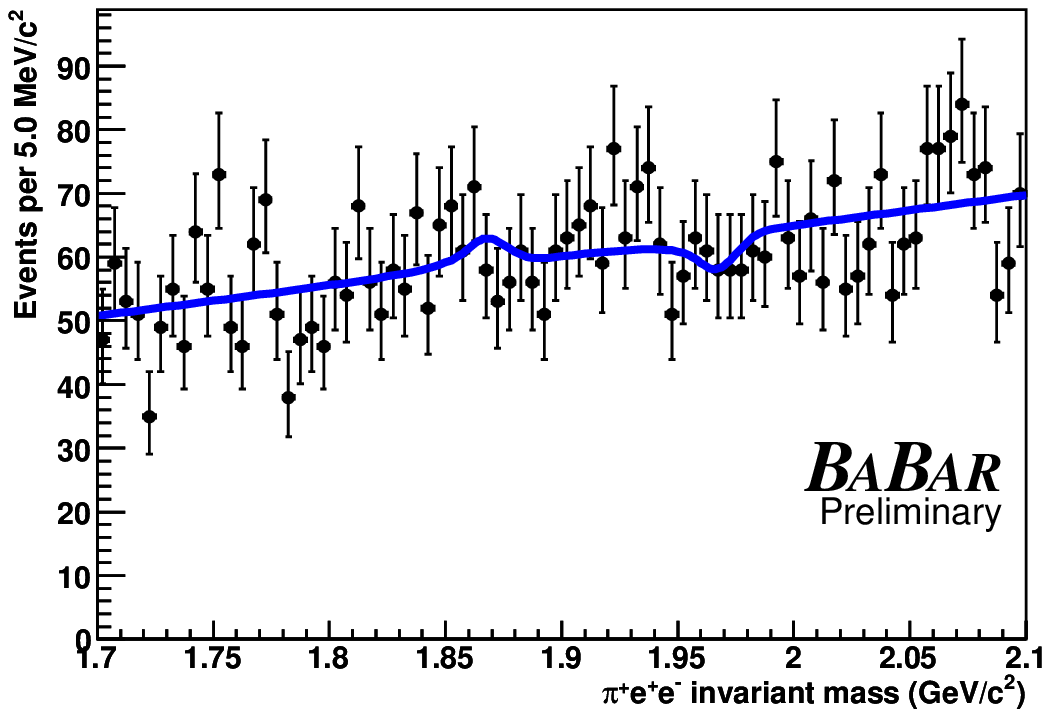}\includegraphics[width=8cm]{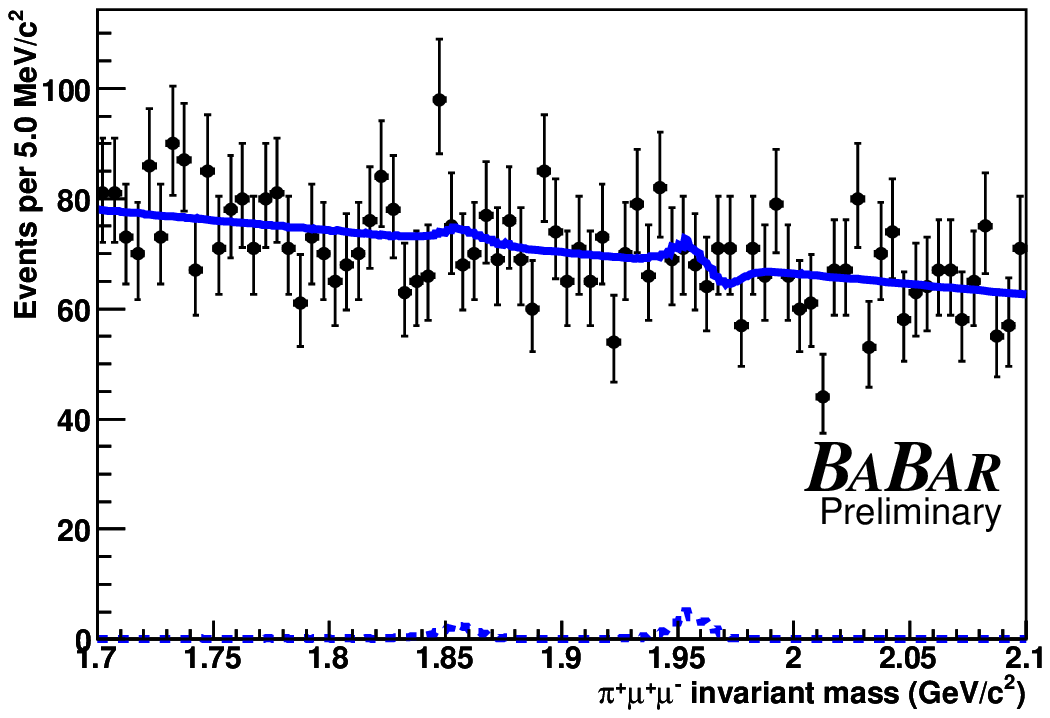}}
\centerline{\includegraphics[width=8cm]{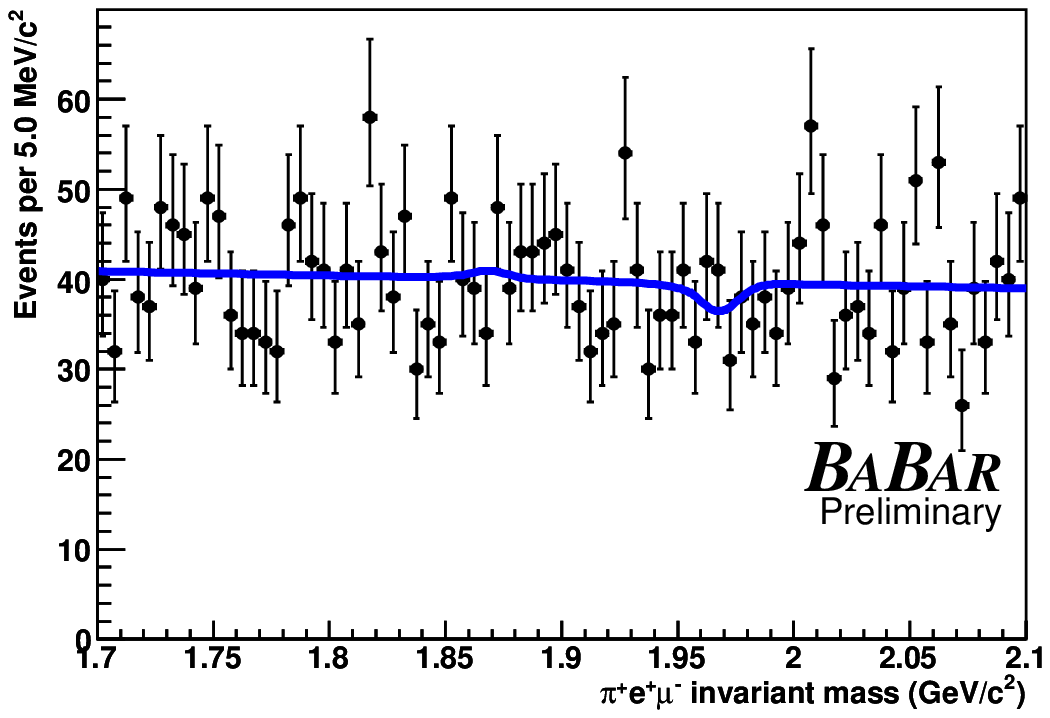}\includegraphics[width=8cm]{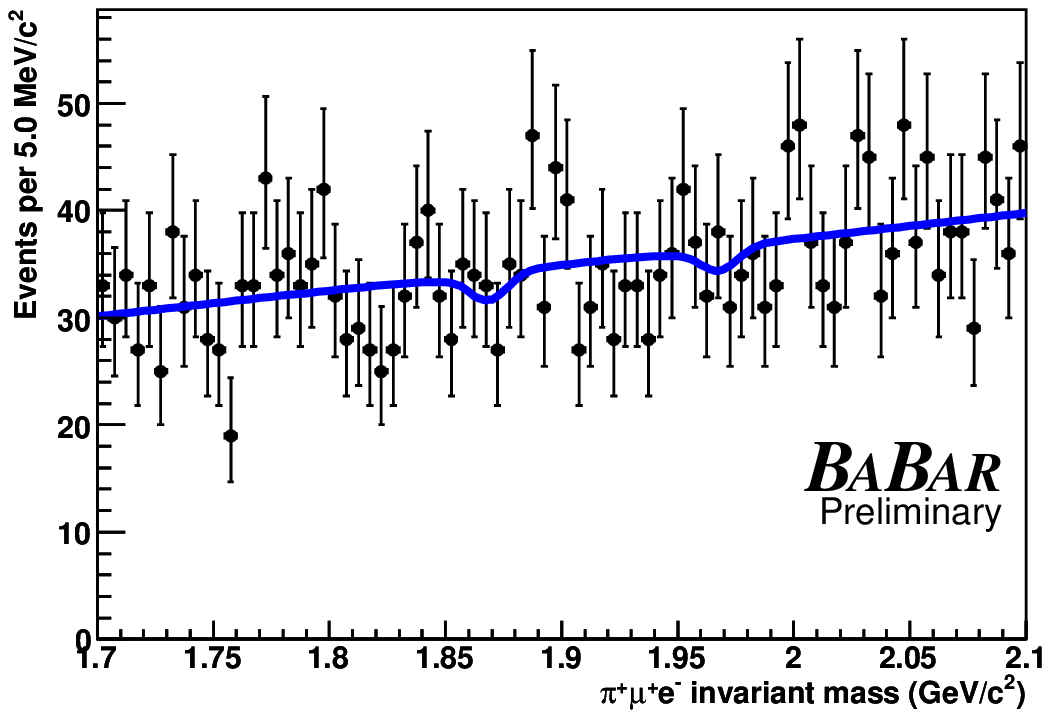}}
\caption{\label{fig:fitDppill} Invariant mass distribution for
$\Dp\to\pip\llp$ candidates.
The solid lines are the results of the fits. The misidentified background component in the dimuon mode is shown as a dashed curve. }
\end{figure}

\begin{figure}[t]
\centerline{\includegraphics[width=8cm]{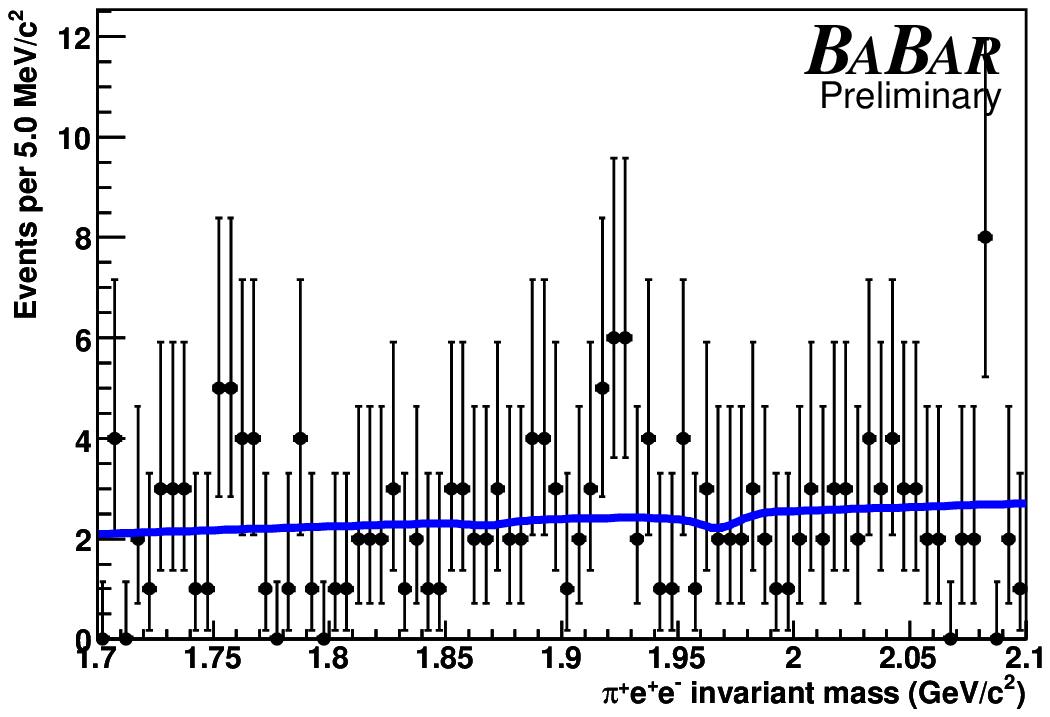}\includegraphics[width=8cm]{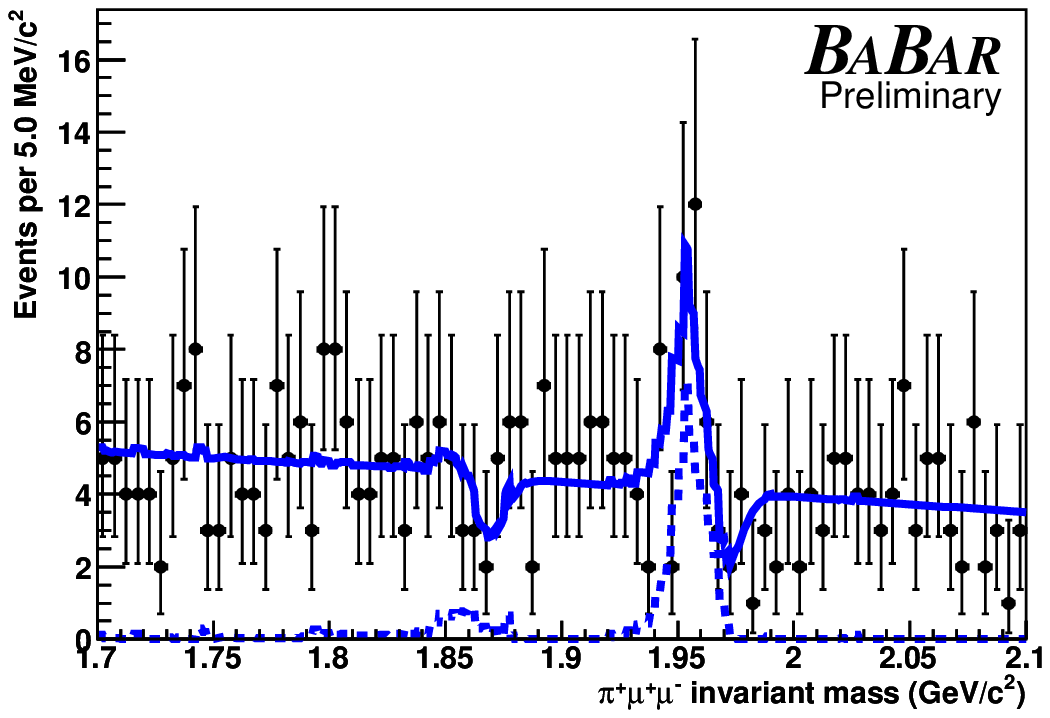}}
\centerline{\includegraphics[width=8cm]{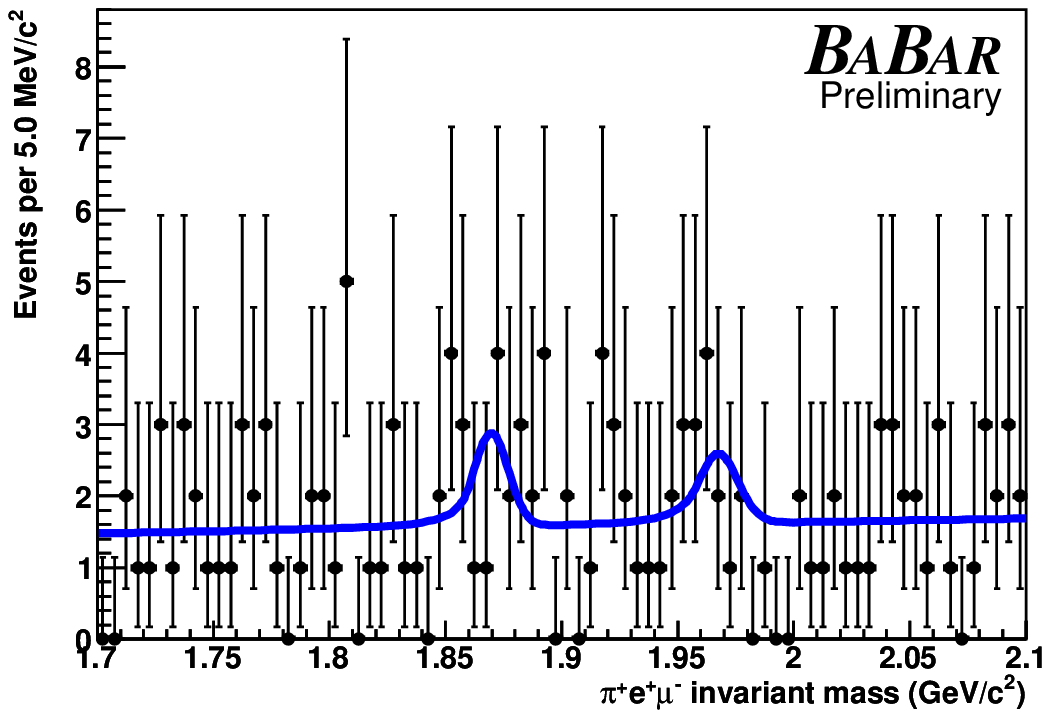}\includegraphics[width=8cm]{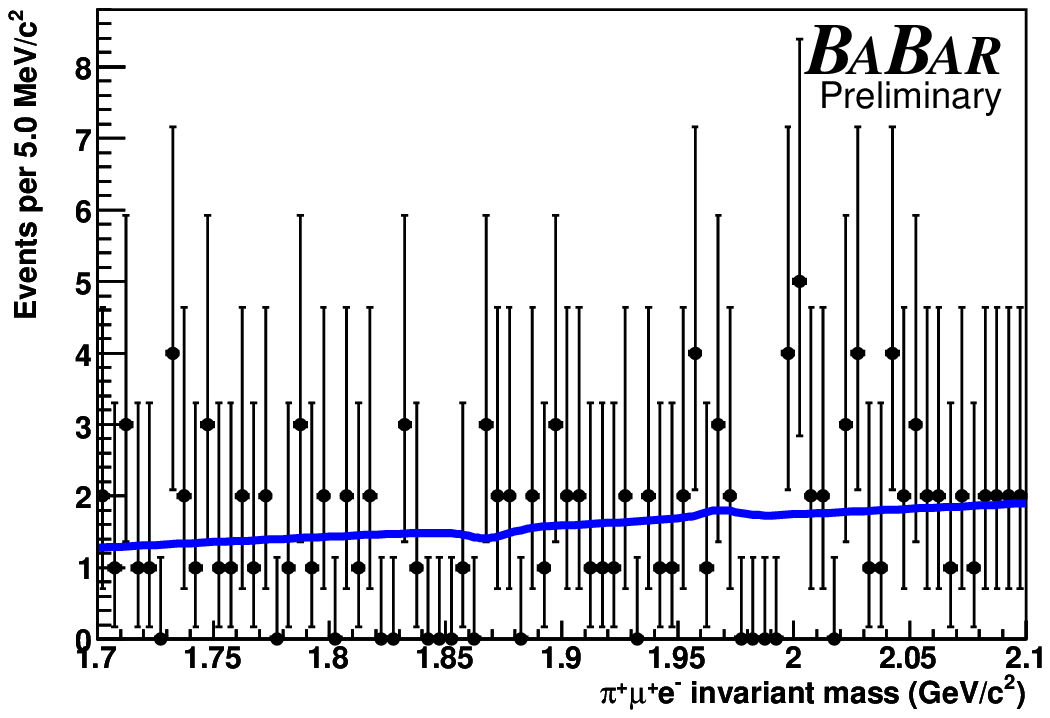}}
\caption{\label{fig:fitDspill} Invariant mass distribution for
$\Ds\to\pip\llp$ candidates.
The solid lines are the results of the fits. The misidentified background component in the dimuon mode is shown as a dashed curve.}
\end{figure}

\begin{figure}[t]
\centerline{\includegraphics[width=8cm]{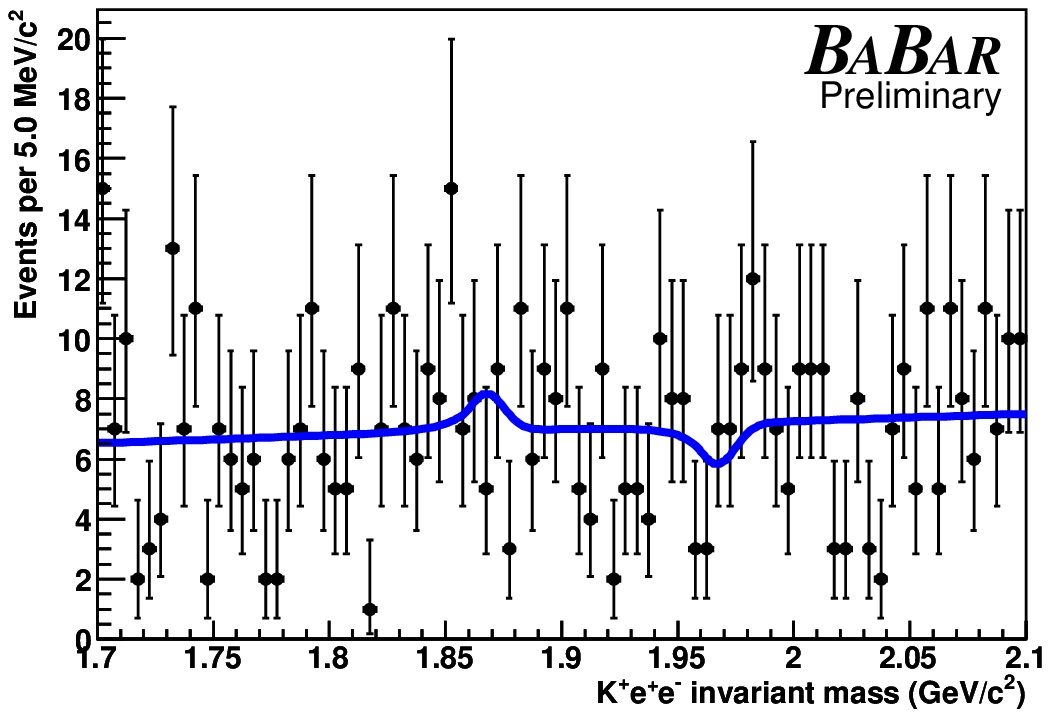}\includegraphics[width=8cm]{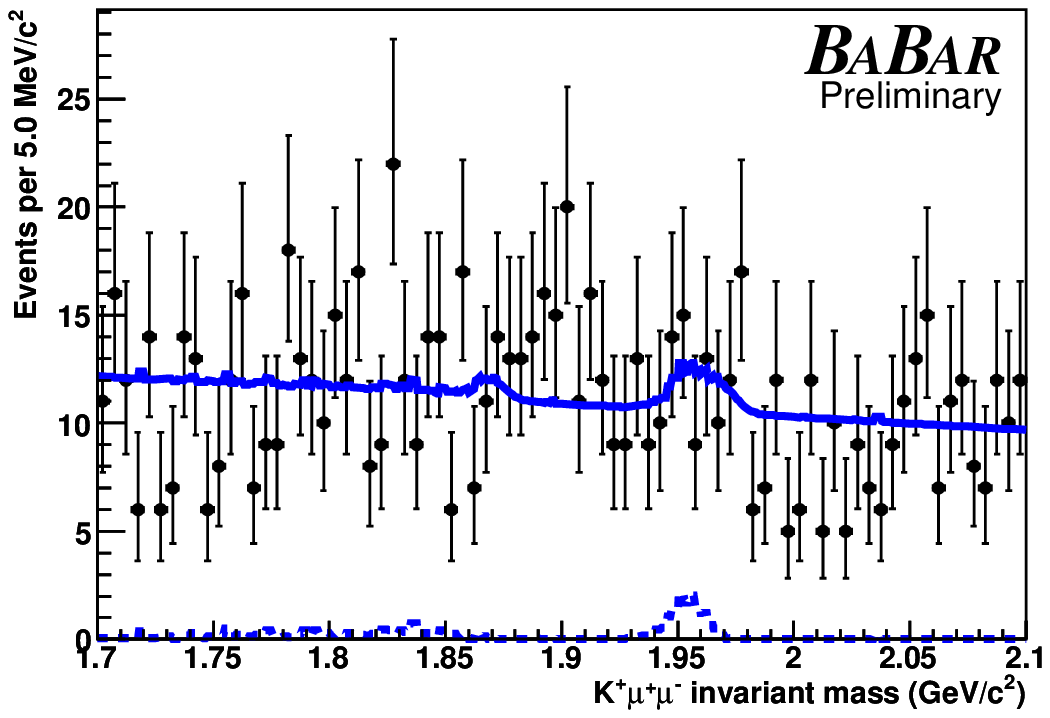}}
\centerline{\includegraphics[width=8cm]{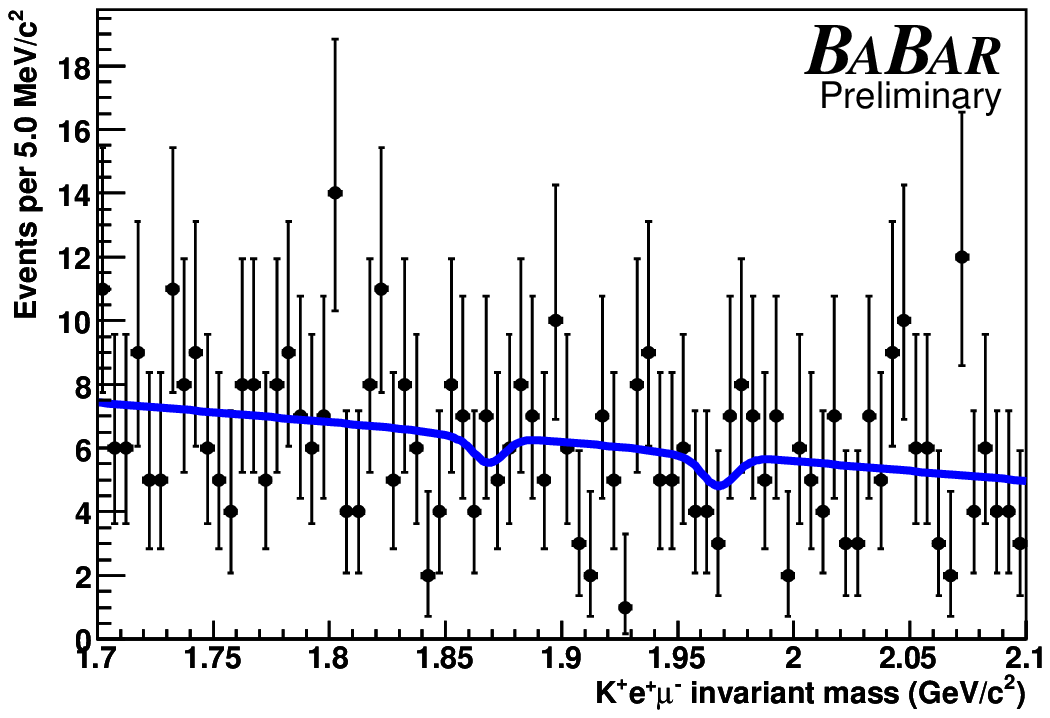}\includegraphics[width=8cm]{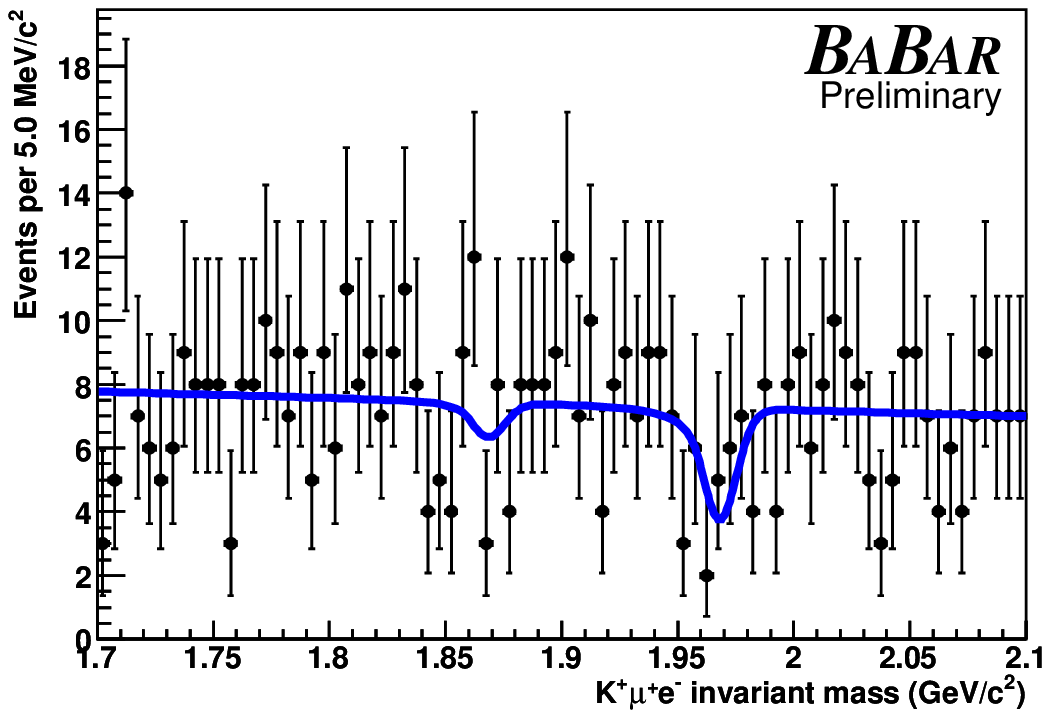}}
\caption{\label{fig:fitDpKll} Invariant mass distribution for
$\Dp\to\Kp\llp$ candidates.
The solid lines are the results of the fits. The misidentified background component in the dimuon mode is shown as a dashed curve.}
\end{figure}

\begin{figure}[t]
\centerline{\includegraphics[width=8cm]{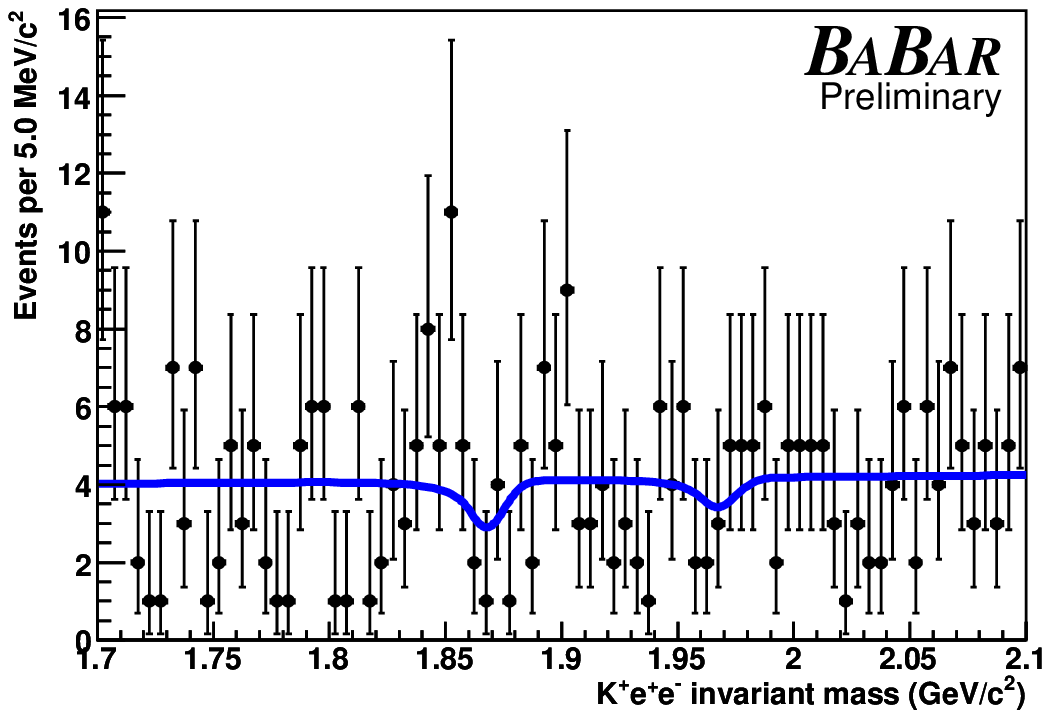}\includegraphics[width=8cm]{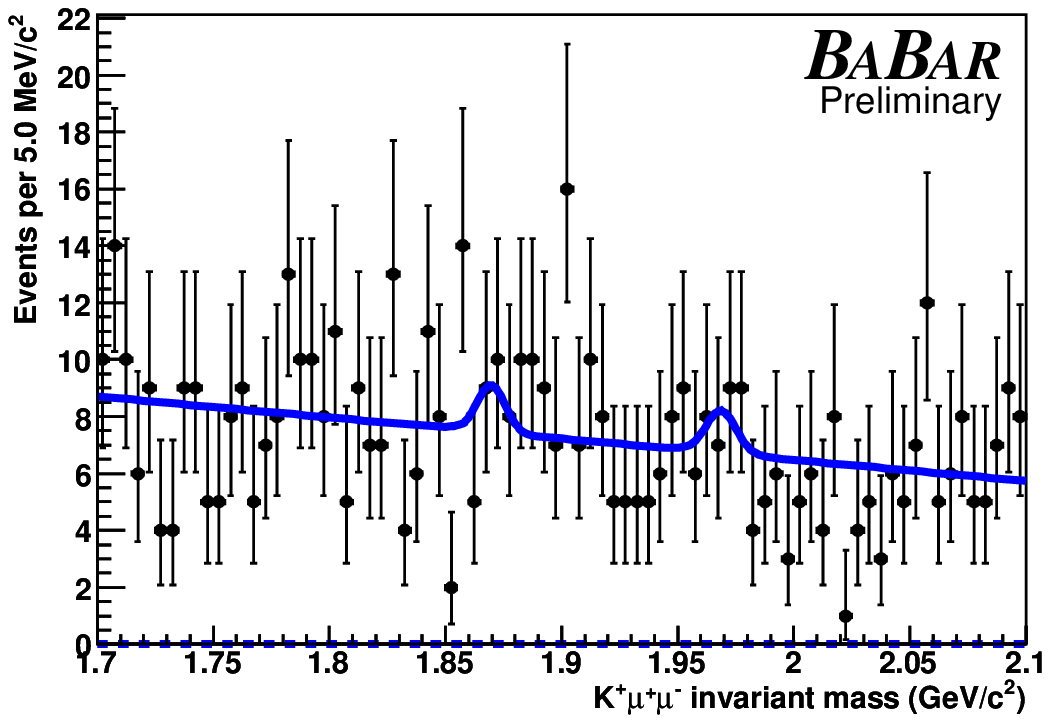}}
\centerline{\includegraphics[width=8cm]{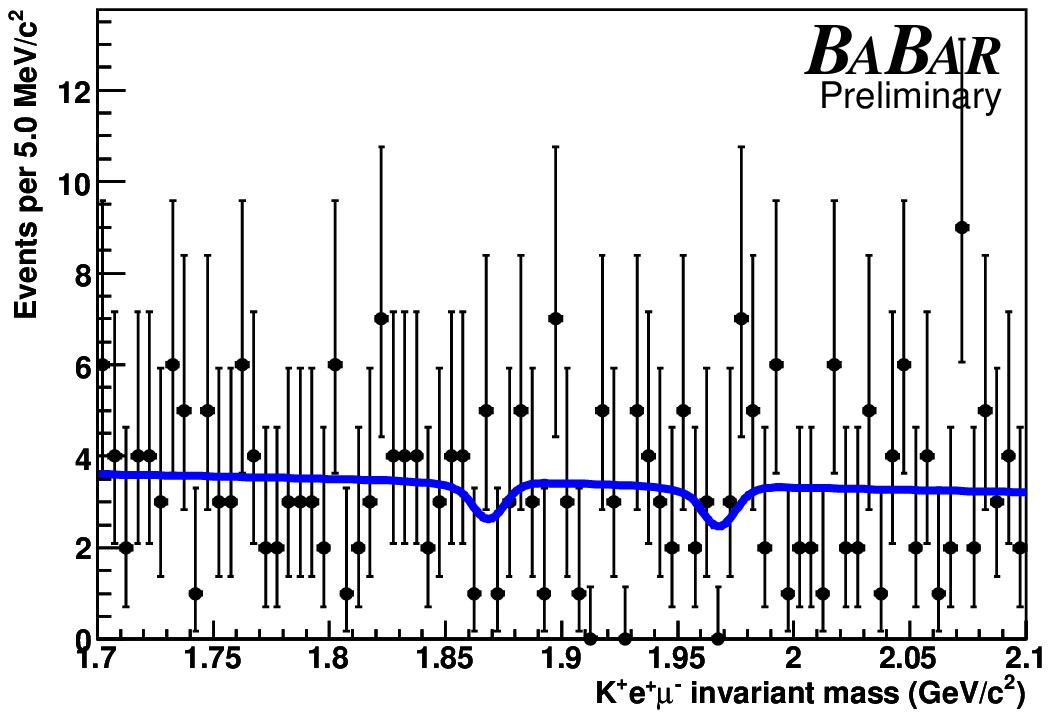}\includegraphics[width=8cm]{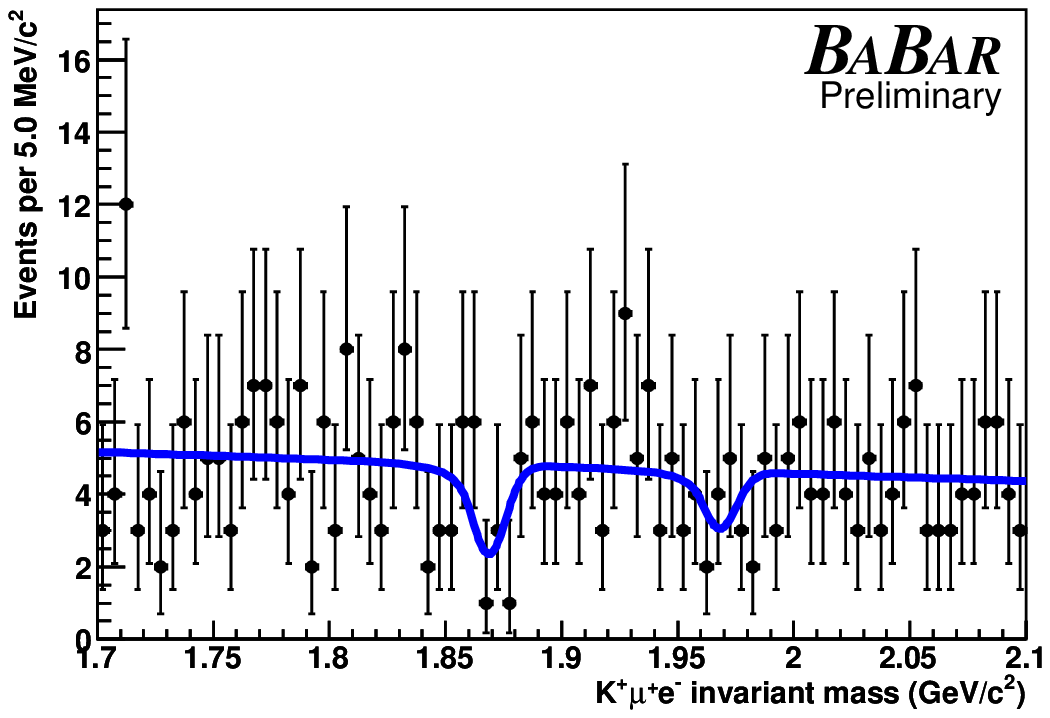}}
\caption{\label{fig:fitDsKll} Invariant mass distribution for
$\Ds\to\Kp\llp$ candidates.
The solid lines are the results of the fits. The misidentified background component in the dimuon mode is shown as a dashed curve.}
\end{figure}

\begin{figure}[t]
\centerline{\includegraphics[width=8cm]{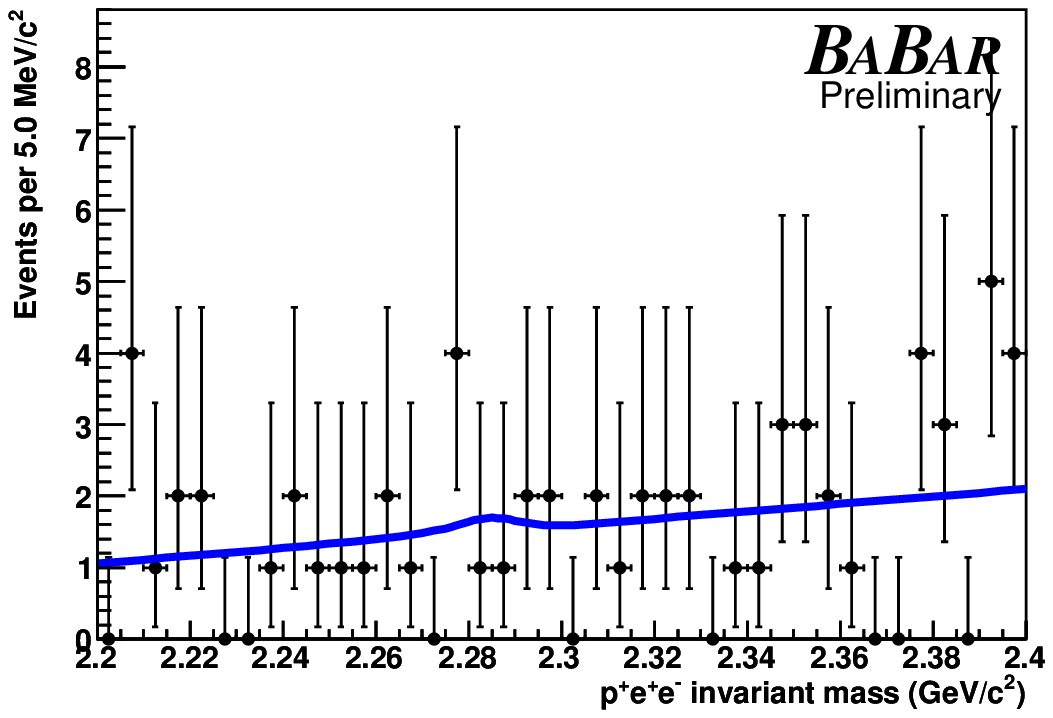}\includegraphics[width=8cm]{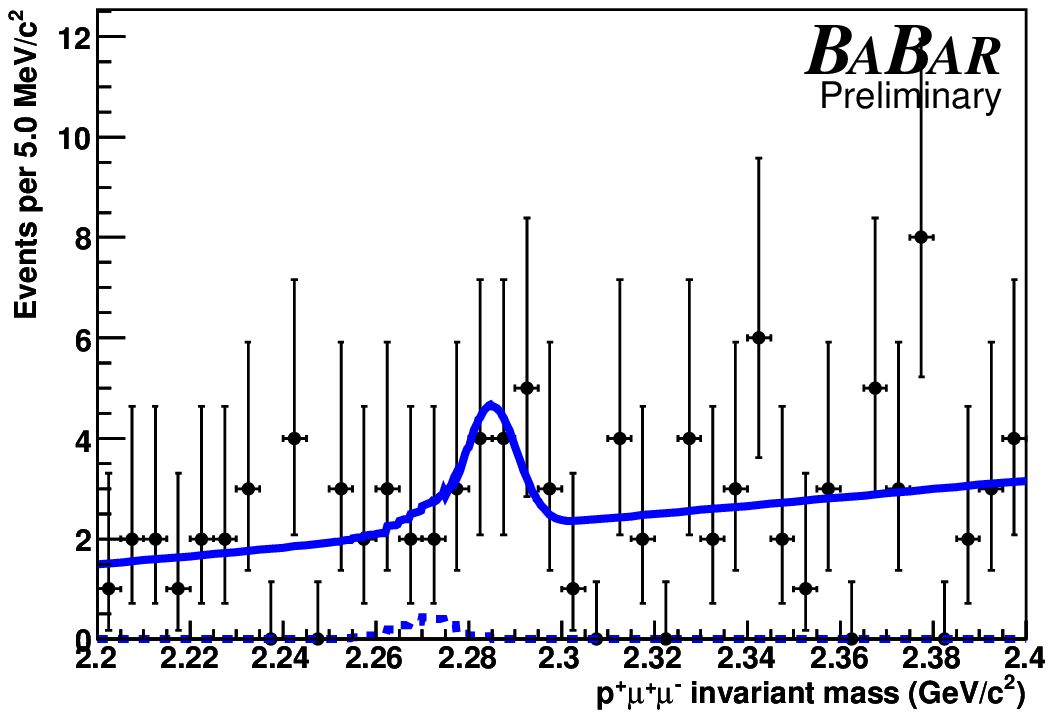}}
\centerline{\includegraphics[width=8cm]{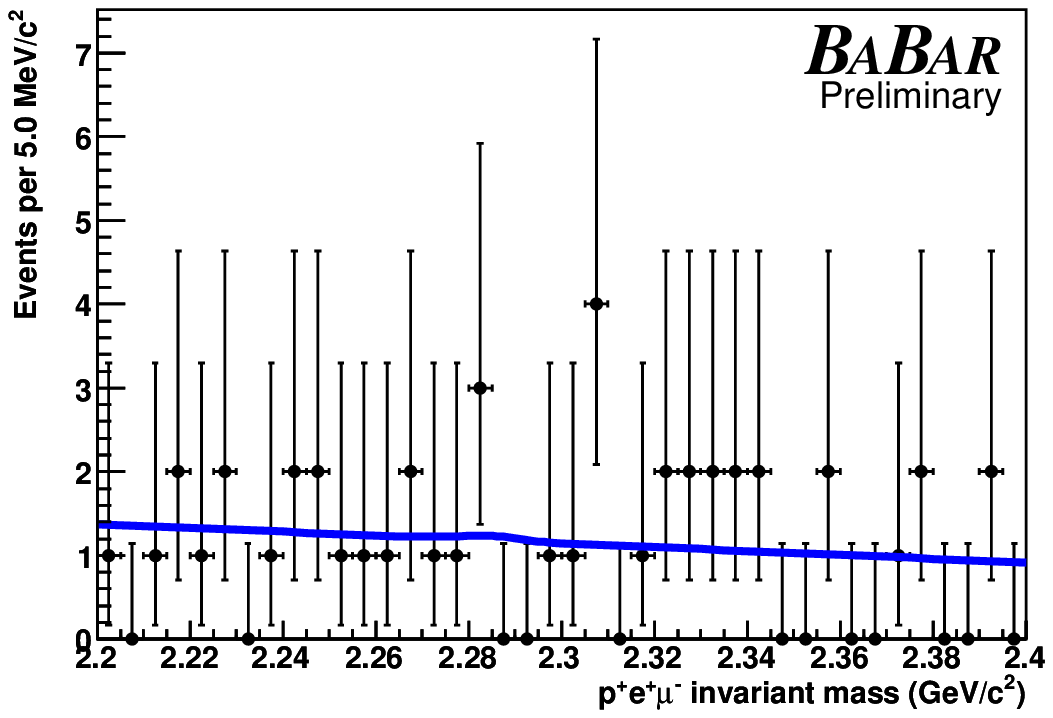}\includegraphics[width=8cm]{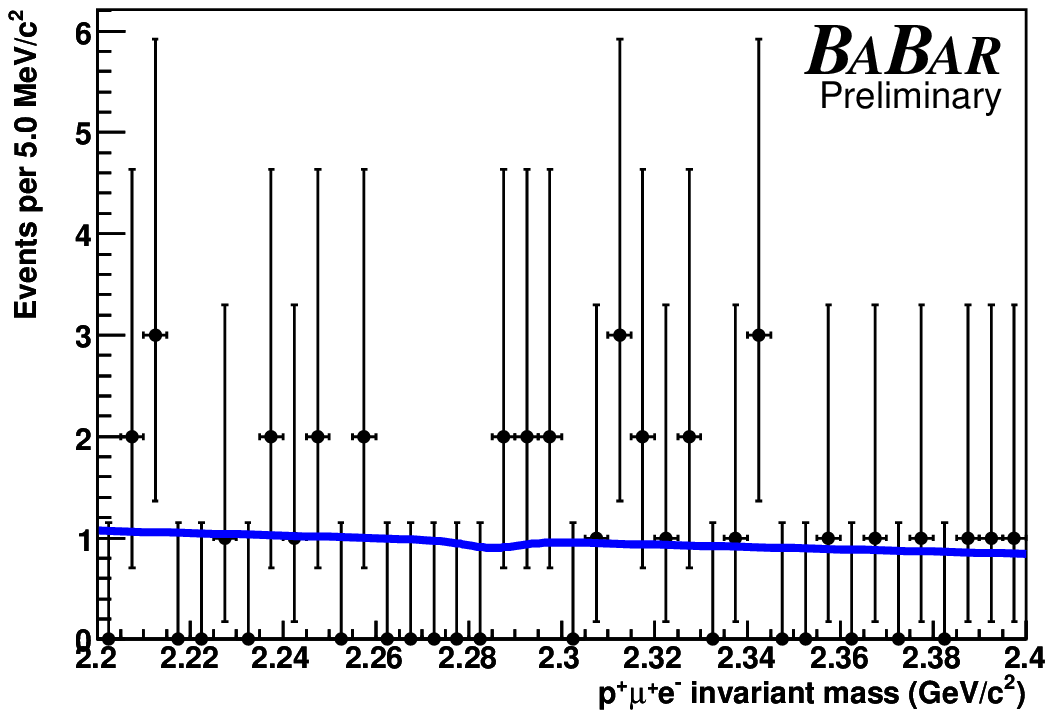}}
\caption{\label{fig:fitLcpll} Invariant mass distribution for
$\Lc\to\proton\llp$ candidates.
The solid lines are the results of the fits. The misidentified background component in the dimuon mode is shown as a dashed curve.}
\end{figure}

The invariant mass distributions for signal candidates in all 20 decay
modes are shown in
Figs.~\ref{fig:fitDppill}--\ref{fig:fitLcpll}. The yields obtained
from unbinned likelihood fits are listed in Table~\ref{tab:fcncFits}
with statistical and systematic uncertainties. Only systematic
uncertainties associated with the signal and background PDFs are
included in the systematic uncertainty for the yields.  The curves
representing the fits are overlaid in the figures. No significant
signals are seen and we calculate upper limits on the branching ratios
$\frac{\Gamma(\Dsp\to\pip\llp)}{\Gamma(\Dsp\to\pip\phi)}$,
$\frac{\Gamma(\Dsp\to\Kp\llp)}{\Gamma(\Dsp\to\pip\phi)}$ and
$\frac{\Gamma(\Lc\to\proton\llp)}{\Gamma(\LctopKpi)}$ at 90\% CL.  For
comparison with previous measurements, the upper limits on the
total branching fraction (BF) at 90\% CL, calculated using
Table~\ref{tab:normBF}, are also given.

The background from misidentified $\Ds\to\pip\pip\pim$ decays in the
\DstoPimm decay channel is found to be $5\pm2$ times larger than
expected from the weighted MC, while in all other muon modes no
significant component of this background is visible. Cross checks of
the MC did not lead to an explanation for the high level of this
background in the \DstoPimm decay channel. Fitting the $\pip\mumu$
candidate mass distribution with the misidentification component fixed
to 0 gives a signal yield of 0.2 events with little change in the
upper limit.

\begin{landscape}
\begin{figure}[t]
\centerline{\includegraphics{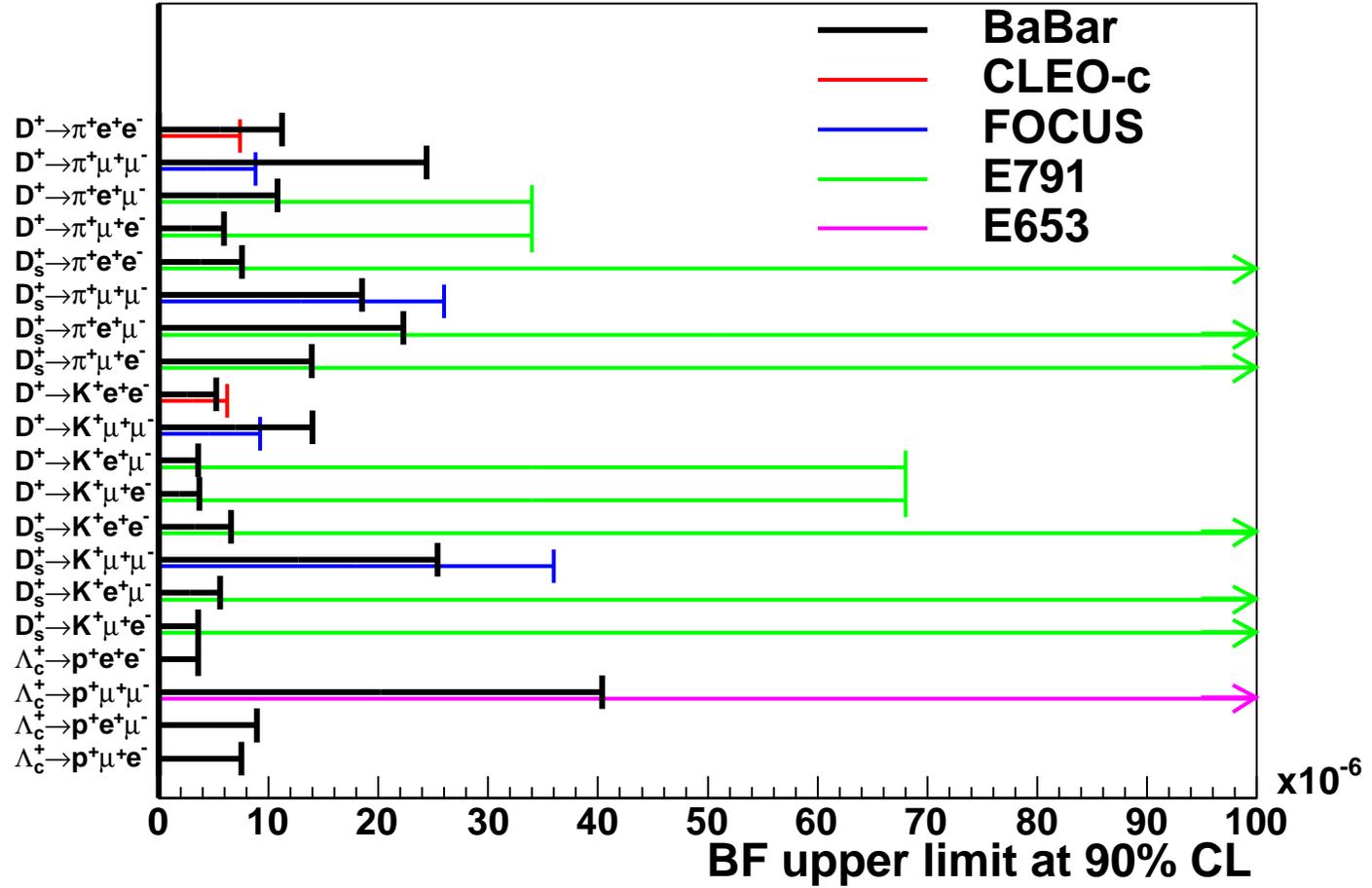}}
\caption{\label{fig:limitCompare} Comparison of the branching fraction limits measured in this
note with previously published measurements \cite{CLEOhee,focus,fixedtarget}.}
\end{figure}
\end{landscape}

Figure~\ref{fig:limitCompare} summarizes the previously published and
the new limits.  For most decay modes, this analysis gives a
significant improvement over the existing measurements.  A recent CLEO
measurement \cite{CLEOhee} is slightly more sensitive to \DtoPiee
decays, and our sensitivity to $\Dp\to\pip\mumu$ and \DtoKmm decays is
worse than fixed target experiments \cite{focus} due to the low muon
efficiency and high backgrounds.  The D$\emptyset$ collaboration has
recently presented a preliminary result \cite{Dzero} that improves
the limit on $\Dp\to\pip\mumu$ decays to $4.7\times 10^{-6}$.

\begin{figure}[t]
\centerline{\includegraphics[width=8cm]{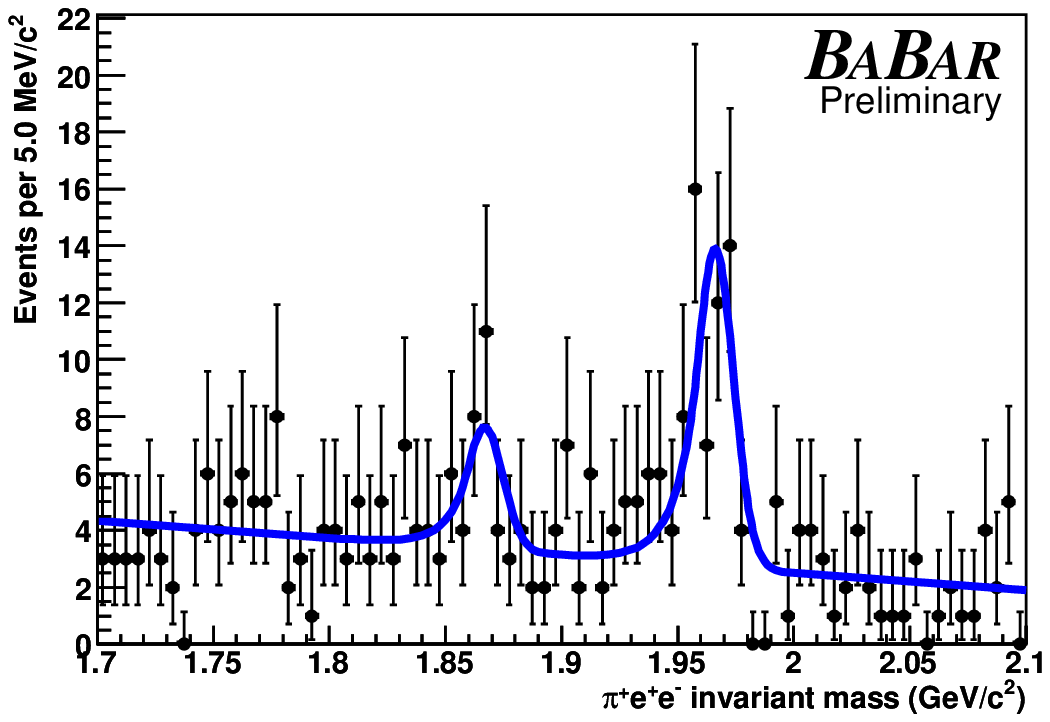}\includegraphics[width=8cm]{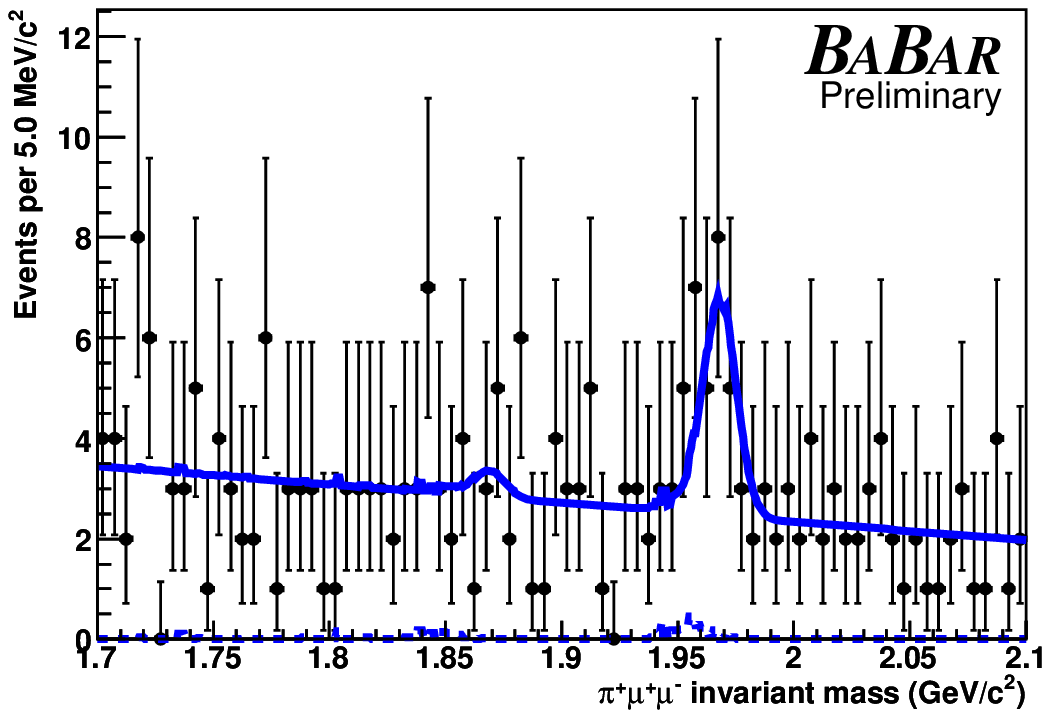}}
\caption{\label{fig:fitDspPiphi} Invariant mass distribution for
(left) $\Dsp\to\pip\phi_{\epem}$ and (right) $\Dsp\to\pip\phi_{\mumu}$
candidates.  The solid lines are the results of the
fits. The misidentified background component in the dimuon mode is
shown as a dashed curve.}
\end{figure}

The $\Dsp\to\pip\phi_{\epem}$ and $\Dsp\to\pip\phi_{\mumu}$ candidates
are shown in Fig.~\ref{fig:fitDspPiphi}. Signals are seen for all decays except
for $\Dp\to\pip\phi_{\mumu}$. Table~\ref{tab:phiModeFits} gives the fit yields
and $\phi\to\ellell$ branching fractions calculated by normalizing to the
\DspTopiphi decay modes. The significance is calculated from the change in likelihood
with and without any signal. The observed branching fractions are in agreement
with the world averages \cite{PDG} of $(2.98\pm0.04)\times 10^{-4}$ for $\phi\to\epem$ and
$(2.85\pm0.19)\times 10^{-4}$ for $\phi\to\mumu$. 

\begin{table}
\caption{\label{tab:fcncFits} Yields from fits to the candidates in the 20
$X_c^+\to h^+\llp$ decay modes.  The first error is statistical and
the second the systematic error on the yield. The third column shows the estimated signal efficiency.
 The fourth column shows the
90\% CL upper limits on the branching ratios of the signal mode to the normalization mode.
The last column shows the limits on the branching fraction for the signal modes at 90\% CL. The upper limits include all systematic uncertainties.
}
\begin{center}
\begin{tabular}{lrrrr}
           & Yield    &            & BR $(10^{-4})$ & BF $(10^{-6})$\\ 
Decay mode & (events) & Efficiency & (90\% CL)  & (90\% CL) \\ \hline
\input tables/fcncResults.tex
\end{tabular}
\end{center}
\end{table}

\begin{table}
\caption{\label{tab:phiModeFits} Yields from fits to the
$\Dsp\to\pip\phi_{\ellell}$ candidates.  The first error is statistical and
the second is the systematic error on the yield. The third column is the
branching fraction for $\phi\to\ellell$ calculated by normalizing to the
$\DspTopiphi$ decay modes. The fourth column is the significance of the signal.}
\begin{center}
\begin{tabular}{lrrrr}
Decay mode & Yield (events) & Efficiency  & BF($\phi\to\ellell$) $(10^{-4})$ & Significance \\ \hline
\input tables/phiModeResults.tex
\end{tabular}
\end{center}
\end{table}

%% file: tables/FullReferenceFits.tex
\DcTopiphi & $\Dp\to\pip/\Kp\llp$       &  25825 p 200 & (4.77$\pm$0.04)\% \\
\DsTopiphi & $\Ds\to\pip\llp$           &  23372 p 162 & (1.26$\pm$0.02)\% \\
\DsTopiphi & $\Ds\to\Kp\llp$            &  71501 p 307 & (3.77$\pm$0.04)\% \\
\LctopKpi  & $\Lc\to\proton\llp$        & 212664 p 1028& (7.38$\pm$0.07)\% \\
\DcTopiphi & $\Dp\to\pip\phi_{\ellell}$ &  32920 p 329 & (6.12$\pm$0.05)\% \\
\DsTopiphi & $\Ds\to\pip\phi_{\ellell}$ & 115480 p 413 & (6.49$\pm$0.05)\% \\ \hline

%% file: tables/fcncResults.tex
\rule{0ex}{2.5ex}$\Dp\to\pip\epem$ & $24.0^{+25.0}_{-24.1}{}^{+ 3.4}_{- 5.1}$      & 3.93\% & $<17.7$    & $<11.2$ \\[0.5ex]
$\Dp\to\pip\mumu$ & $ 1.5^{+20.1}_{-19.3}{}^{+ 3.4}_{- 2.6}$      & 1.09\% & $<38.7$    & $<24.4$ \\[0.5ex]
$\Dp\to\pip\ep\mun$ & $ 4.1^{+17.8}_{-16.3}{}^{+ 3.1}_{- 2.1}$    & 2.27\% & $<17.1$  & $<10.8$ \\[0.5ex]
$\Dp\to\pip\mup\en$ & $-12.1^{+15.5}_{-14.8}{}^{+ 3.2}_{- 0.0}$   & 2.29\% & $< 9.3$ & $< 5.9$ \\[0.5ex]
$\Ds\to\pip\epem$ & $-1.7^{+ 5.3}_{- 4.6}{}^{+ 0.2}_{- 2.0}$      & 1.14\% & $< 2.1$    & $< 7.6$ \\[0.5ex]
$\Ds\to\pip\mumu$ & $-9.4^{+ 5.0}_{- 4.4}{}^{+ 0.2}_{- 1.4}$      & 0.31\% & $< 5.1$    & $<18.5$ \\[0.5ex]
$\Ds\to\pip\ep\mun$ & $ 4.8^{+ 4.7}_{- 3.9}{}^{+ 0.8}_{- 0.3}$    & 0.66\% & $< 6.2$  & $<22.3$ \\[0.5ex]
$\Ds\to\pip\mup\en$ & $ 0.5^{+ 4.0}_{- 3.3}{}^{+ 1.0}_{- 0.1}$    & 0.65\% & $< 3.8$  & $<13.9$ \\[0.5ex]
$\Dp\to\Kp\epem$ & $ 5.9^{+ 8.9}_{- 7.8}{}^{+ 3.8}_{- 0.3}$       & 3.21\% & $< 8.2$     & $< 5.2$ \\[0.5ex]
$\Dp\to\Kp\mumu$ & $ 2.9^{+ 8.0}_{- 7.0}{}^{+ 0.2}_{- 3.7}$       & 0.75\% & $<22.2$     & $<14.0$ \\[0.5ex]
$\Dp\to\Kp\ep\mun$ & $-3.4^{+ 6.5}_{- 5.6}{}^{+ 1.0}_{- 0.1}$     & 1.64\% & $< 5.7$   & $< 3.6$ \\[0.5ex]
$\Dp\to\Kp\mup\en$ & $-4.4^{+ 7.1}_{- 6.1}{}^{+ 1.4}_{- 3.0}$     & 1.64\% & $< 5.9$   & $< 3.7$ \\[0.5ex]
$\Ds\to\Kp\epem$ & $-3.8^{+ 6.2}_{- 5.3}{}^{+ 1.5}_{- 1.3}$       & 2.81\% & $< 1.8$     & $< 6.6$ \\[0.5ex]
$\Ds\to\Kp\mumu$ & $ 5.0^{+ 6.5}_{- 6.1}{}^{+ 0.1}_{- 0.3}$       & 0.68\% & $< 7.1$     & $<25.4$ \\[0.5ex]
$\Ds\to\Kp\ep\mun$ & $-3.7^{+ 5.1}_{- 4.4}{}^{+ 1.4}_{- 1.4}$     & 1.40\% & $< 1.5$   & $< 5.6$ \\[0.5ex]
$\Ds\to\Kp\mup\en$ & $-6.5^{+ 4.9}_{- 4.3}{}^{+ 0.2}_{- 1.1}$     & 1.40\% & $< 1.0$   & $< 3.6$ \\[0.5ex]
$\Lc\to\proton\epem$ & $ 0.9^{+ 4.1}_{- 3.4}{}^{+ 0.4}_{- 0.1}$   & 4.11\% & $< 0.7$  & $< 3.6$ \\[0.5ex]
$\Lc\to\proton\mumu$ & $ 6.9^{+ 4.7}_{- 3.7}{}^{+ 0.3}_{- 0.6}$   & 0.67\% & $< 8.1$  & $<40.4$ \\[0.5ex]
$\Lc\to\proton\ep\mun$ & $ 0.2^{+ 2.9}_{- 2.0}{}^{+ 0.5}_{- 0.5}$ & 1.19\% & $< 1.8$ & $< 8.9$ \\[0.5ex]
$\Lc\to\proton\mup\en$ & $-0.2^{+ 2.5}_{- 1.7}{}^{+ 0.5}_{- 0.9}$ & 1.18\% & $< 1.5$ & $< 7.5$ \\ \hline

%% file: tables/phiModeResults.tex
\rule{0ex}{2.5ex}$\Dp\to\pip\phi_{\epem}$ & $19.0^{+ 7.1}_{- 5.9}{}^{+ 0.4}_{- 2.9}$ & 4.11\% & $ 3.9^{+ 1.5}_{- 1.2}{}^{+ 0.2}_{- 0.6}$ & $2.7\sigma$ \\[0.5ex]
$\Dp\to\pip\phi_{\mumu}$ & $ 1.8^{+ 4.5}_{- 3.7}{}^{+ 0.1}_{- 0.7}$ & 0.67\% & $< 8.2$ at 90\% CL & $0.3\sigma$ \\[0.5ex]
$\Ds\to\pip\phi_{\epem}$ & $50.5^{+ 9.0}_{- 8.1}{}^{+ 1.6}_{- 3.9}$ & 1.19\% & $ 2.8^{+ 0.5}_{- 0.5}{}^{+ 0.2}_{- 0.3}$ & $6.8\sigma$ \\[0.5ex]
$\Ds\to\pip\phi_{\mumu}$ & $15.3^{+ 6.1}_{- 4.8}{}^{+ 1.3}_{- 0.3}$ & 1.18\% & $ 3.5^{+ 1.4}_{- 1.1}{}^{+ 0.4}_{- 0.2}$ & $3.2\sigma$ \\ \hline

%% file: result.tex
\section{SUMMARY}

A search for the decay modes $\Dsp\to\pip\llp$, $\Dsp\to\Kp\llp$ and
$\Lc\to\proton\llp$ has been performed using 288\invfb of \epem\
data. No signals are observed and we obtain upper limits on the
branching ratios
$\frac{\Gamma(\Dsp\to\pip\llp)}{\Gamma(\Dsp\to\pip\phi)}$,
$\frac{\Gamma(\Dsp\to\Kp\llp)}{\Gamma(\Dsp\to\pip\phi)}$ and
$\frac{\Gamma(\Lc\to\proton\llp)}{\Gamma(\LctopKpi)}$ between
$10^{-4}$ and $40\times10^{-4}$ at 90\% CL. This corresponds to limits
on the branching fractions between $4\times 10^{-6}$ and $4\times
10^{-5}$.  These limits are calculated under the assumption of
three-body phase-space decays; the efficiency varies by up to 25\% as
a function of dilepton invariant mass. For 17 out the 20 decay modes,
the limits are an improvement over the existing measurements.

%% file: pubboard/acknowledgements.tex
We are grateful for the 
extraordinary contributions of our \pep2\ colleagues in
achieving the excellent luminosity and machine conditions
that have made this work possible.
The success of this project also relies critically on the 
expertise and dedication of the computing organizations that 
support \babar.
The collaborating institutions wish to thank 
SLAC for its support and the kind hospitality extended to them. 
This work is supported by the
US Department of Energy
and National Science Foundation, the
Natural Sciences and Engineering Research Council (Canada),
Institute of High Energy Physics (China), the
Commissariat \`a l'Energie Atomique and
Institut National de Physique Nucl\'eaire et de Physique des Particules
(France), the
Bundesministerium f\"ur Bildung und Forschung and
Deutsche Forschungsgemeinschaft
(Germany), the
Istituto Nazionale di Fisica Nucleare (Italy),
the Foundation for Fundamental Research on Matter (The Netherlands),
the Research Council of Norway, the
Ministry of Science and Technology of the Russian Federation, and the
Particle Physics and Astronomy Research Council (United Kingdom). 
Individuals have received support from 
the Marie-Curie IEF program (European Union) and
the A. P. Sloan Foundation.